\documentclass[letterpaper,11pt]{article}
\pdfoutput=1
\usepackage{jhepmod} 

\usepackage{booktabs}
\usepackage[english]{babel}
\usepackage{amsmath,amssymb,amsbsy,amstext, amsthm, simplewick, amsfonts}
\usepackage{hyperref}
\usepackage{graphicx}
\usepackage{subcaption}

\usepackage{siunitx}
\usepackage{upgreek}
\usepackage{framed}
\usepackage{wrapfig}
\usepackage{multirow}
\usepackage{bbm}
\usepackage[svgnames,dvipsnames,x11names]{xcolor}
\usepackage[utf8x]{inputenc}
\usepackage{selinput}
\usepackage{bm}
\usepackage{float}
\usepackage{yfonts}
\usepackage{mathtools}
\usepackage{enumitem}   
\usepackage{longtable}
\usepackage{tabularx}

\usepackage{tikz}
\usetikzlibrary{arrows,chains,matrix,positioning,scopes}
\usepackage{tikz-cd}
\usetikzlibrary{matrix, calc, arrows}

\def\sraise{\;\raise1.0pt\hbox{$'$}\hskip-6pt\partial}
\def\slower{\;\overline{\raise1.0pt\hbox{$'$}\hskip-6pt\partial}}
\def\tline{\specialrule{0.7pt}{0pt}{0pt}}

\usepackage{colortbl}

\makeatletter
\newlength{\apb@width}
\newcommand{\autoparbox}[2][c]{\settowidth{\apb@width}{#2}\parbox[#1]{\apb@width}{#2}}

\makeatother

\def\d{{\rm d}}

\def\k{{\boldsymbol k}}

\def\n{{\boldsymbol n}}
\def\q{{\boldsymbol q}}

\def\r{{\boldsymbol r}}
\def\s{{\boldsymbol s}}
\def\t{{\boldsymbol t}}
\def\u{{\boldsymbol u}}
\def\v{{\boldsymbol v}}

\def\x{{\boldsymbol x}}

\def\0{{\boldsymbol{0}}}

\def\D{{\cal D}}

\def\H{{\cal H}}
\def\G{{\cal G}}

\def\I{{\sf I}}

\def\M{{\cal M}}

\def\O{{\cal O}}

\def\W{{\sf W}}

\def\nn{\nonumber\\}

\def\Ddelta{\delta_{\rm D}}

\def\fnl{f_{\rm NL}}
\def\tnl{\tau_{\rm NL}}
\def\gnl{g_{\rm NL}}

\def\hs{\hskip 1pt}

\def\beq{\begin{equation}}
\def\eeq{\end{equation}}
\def\be{\begin{equation}}
\def\ee{\end{equation}}

\usepackage[framemethod=default]{mdframed}
\newmdenv[skipabove=7pt,
skipbelow=7pt,
rightline=false,
leftline=false,
topline=false,
bottomline=false,
backgroundcolor=gray!10,
linecolor=gray,
innerleftmargin=5pt,
innerrightmargin=5pt,
innertopmargin=5pt,
innerbottommargin=5pt,
leftmargin=0cm,
rightmargin=0cm,
linewidth=4pt]{eBox}

\title{Cosmological Angular Trispectra and Non-Gaussian Covariance}

\author{Hayden Lee and Cora Dvorkin}
\affiliation{Department of Physics, Harvard University, 17 Oxford Street, Cambridge, MA 02138, USA}\emailAdd{hylee@g.harvard.edu, cdvorkin@g.harvard.edu}

\abstract{
Angular cosmological correlators are infamously difficult to compute due to the highly oscillatory nature of the projection integrals. Motivated by recent development on analytic approaches to cosmological perturbation theory, in this paper we present an efficient method for computing cosmological four-point correlations in angular space, generalizing previous works on lower-point functions. This builds on the FFTLog algorithm that approximates the matter power spectrum as a sum over power-law functions, which makes certain momentum integrals analytically solvable.
The computational complexity is drastically reduced for correlators in a ``separable'' form---we define a suitable notion of separability for cosmological trispectra, and derive formulas for angular correlators of different separability classes.
As an application of our formalism, we compute the angular galaxy trispectrum at tree level, with and without primordial non-Gaussianity. This includes effects of redshift space distortion and bias parameters up to cubic order. We also compute the non-Gaussian covariance of the angular matter power spectrum due to the connected four-point function, beyond the Limber approximation. We demonstrate that, in contrast to the standard lore, the Limber approximation can fail for the non-Gaussian covariance computation even for large multipoles.}

\begin{document}

\maketitle
\flushbottom

\newpage

\section{Introduction}

Our knowledge of the composition of the universe and the physics of its early stages has undergone tremendous advances over the last three decades. So far, the best constraints on the statistics of the primordial fluctuations are provided by detailed measurements of the cosmic microwave background (CMB)~\cite{Planck2018}, which will soon be improved with the advent of the next generation CMB experiments, measuring polarization~\cite{Abazajian:2019eic,Ade:2018sbj,CLASS} and temperature fluctuations at small scales~\cite{Wu:2019hek,Louis:2016ahn} with higher precision.  
Complementary to the CMB, upcoming surveys such as the Large Synoptic Survey Telescope (LSST)~\cite{Abate:2012za}, SPHEREx~\cite{Dore:2014cca}, Euclid~\cite{Euclid}, and the Dark Energy Spectroscopic Instrument (DESI)~\cite{Aghamousa:2016zmz} will be refining the measurements of the large-scale structure (LSS) of the universe, using different probes involving galaxy clustering, weak lensing, 21-cm emission line, etc. 
Containing three-dimensional information, these LSS surveys can ultimately surpass the CMB in providing stronger constraints on the statistics of the scalar fluctuations, and thus offering a better window into primordial non-Gaussianity.

\vskip 4pt 
Understanding the LSS is inherently more difficult than the CMB due to the intrinsically nonlinear nature of gravity.
The gravitational evolution of dark matter density is most commonly studied in Fourier space, either using the theoretical framework of cosmological perturbation theory (and its extensions) or through numerical $N$-body simulations. 
In practice, we do not directly observe distributions of dark matter, but rather that of their tracers such as halos and galaxies. The relationship between dark matter and its tracers is called {\it bias}, and provides a crucial link between observations and the physics of cosmological perturbations. Amongst the two biased tracers, the fact that halos are nothing but gravitationally bound matter makes it possible for us simulate them in gravity-only simulations, and empirically study their bias. In contrast, galaxy dynamics is vastly more complicated, and currently we do not have first principles understanding of its formation process. This makes it difficult to reliably make theoretical predictions for galaxy distributions or simulating them on cosmological scales. 

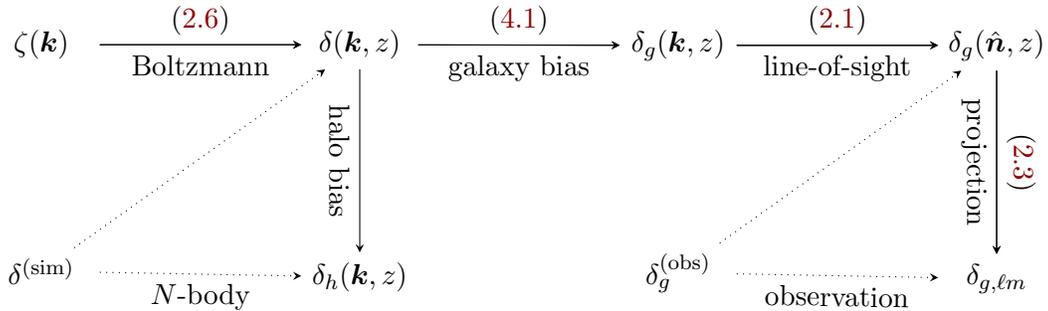
\begin{figure}    
\centering
\begin{tikzpicture}
  \matrix (m) [matrix of math nodes,row sep=6em,column sep=7em,minimum width=4em]
   { \zeta(\k) &  \delta(\k,z) & \delta_g(\k,z)  & \delta_g(\hat\n,z)   \\
      \delta^{(\rm sim)} & \delta_h(\k,z) & \delta_g^{(\rm obs)} & \delta_{g, \ell m}  \\
 };
  \path[-stealth]
  (m-1-1) edge node [above] { \eqref{transfer}} (m-1-2) 
    (m-1-1) edge node [below] {Boltzmann} (m-1-2) 
  	(m-1-2) edge node [above] {\eqref{eq:bias} } (m-1-3)
    (m-1-2) edge node [below,sloped] {halo bias} (m-2-2)
    (m-1-3) edge node [above] {\eqref{Olm}} (m-1-4)
        (m-1-3) edge node [below] {line-of-sight} (m-1-4)
        (m-1-2) edge node [below] {galaxy bias} (m-1-3)
    (m-1-4) edge node [above,sloped] {\eqref{Olm2}} (m-2-4)
        (m-1-4) edge node [below,sloped] {projection} (m-2-4)
	(m-2-1) edge [dotted] node [below] {$N$-body} (m-2-2)
	(m-2-1) edge [dotted] node [above,sloped] {} (m-1-2)
		(m-2-3) edge [dotted] node [above,sloped] {} (m-1-4)
			(m-2-3) edge [dotted] node [below,sloped] {observation} (m-2-4)
  ;
\end{tikzpicture}
\caption{Relations between scalar perturbations in cosmology. The initial curvature perturbation $\zeta$ is evolved at late times into the matter density field $\delta$, which gives rise to its tracer density fields $\delta_h$ and $\delta_g$ for halos and galaxies, respectively. The matter and halo densities can be simulated, while the galaxy densities are observed directly in redshift space.}
\label{fig:scalar}
\end{figure}

\vskip 4pt
At present, direct observations provide the best means of studying galaxy distributions. 
When surveys have limited sky coverage, the sky is effectively flat, which makes it possible to do a Fourier analysis. Near-future spectroscopic redshift surveys will become deeper and wider, which instead requires a full-sky formalism of computing cosmological observables.
Conventionally, it is still preferred to do cosmological analyses in Fourier space, in which theoretical calculations can be most naturally done. This, however, involves one extra complication: we must assume a fiducial cosmology to translate redshifts into distances, which, if differs from the true cosmology, results in distortions of data.\footnote{This simply follows from the fact that the comoving distance between two objects with different redshifts is given by an integral involving the cosmology-dependent Hubble parameter over the redshift difference. Accounting for this requires a modeling of the so-called Alcock-Paczynski effect~\cite{Alcock:1979mp}.} 
No such assumption is needed when working in redshift space, since observations directly map the positions of galaxies in terms of their redshifts and angles. Angular correlation functions in redshift space are therefore in principle the most natural observable to describe galaxy distributions (see Fig.~\ref{fig:scalar}). It is therefore desirable to be able to directly compare our theoretical models with observations in redshift space.

\vskip 4pt
There are, however, well-known numerical challenges involved in projecting cosmological observables onto two-dimensional redshift surfaces. 
The main difficulty is that the projection integrals consist of products of spherical Bessel functions that are highly oscillatory, requiring a large number of integration points to reach a desired accuracy.  
The computational cost quickly becomes infeasible as we go to higher points due to the shear multi-dimensionality of integrals. 
Worse, cosmological parameter estimation using Markov chain Monte Carlo methods in upcoming galaxy surveys would require observables to be computed millions of times, summed over many redshift bins as well as their cross-correlations.
All these limitations have so far restricted performing cosmological data analyses in redshift space far from ideal.  

\vskip 4pt
Over the past few years, there have emerged novel approaches to overcome the numerical obstacles of computing angular observables in cosmology. One of the most insightful ideas has been the use of the so-called FFTLog algorithm, originally proposed in~\cite{Hamilton:1999uv}. In this approach, one decomposes the matter power spectrum into a sum of power-law functions, in attempts to analytically solve certain integrals that are otherwise difficult to evaluate numerically. It turns out that this analytic method not only leads to a dramatic increase in the speed of the calculations, but also significantly improves the numerical stability of the integrals. The utility of this method, and similar methods that bypass the spherical Bessel integrals, has been explored in a number of recent works~\cite{Assassi:2017lea, Schoneberg:2018fis, Simonovic:2017mhp, Gebhardt:2017chz, Slepian:2018vds, Fang:2019xat, Slepian:2019wia}. 
These theoretical studies on angular correlators, and codes for computing them (e.g.~\cite{Campagne:2017xps, Campagne:2017wec, Tansella:2017rpi, Tansella:2018sld, Gebhardt:2017chz}), have so far restricted their analyses to two-point and three-point statistics, due to yet existing complications of higher-point functions.

\vskip 4pt
However, there are good motivations to go beyond and study four-point statistics. 
First of all, shapes of four-point functions can exhibit new qualitative features that provide useful information for constraining the physics of inflation beyond what is available in three-point functions. For instance, many inflationary models involving light degrees of freedom produce four-point functions whose sizes can be larger than that of three-point functions.
Also, the trispectrum---the Fourier counterpart of the four-point function---of biased tracers from gravitational evolution at tree level has contributions from bias parameters up to third order, which also enter the power spectrum calculation at two-loop. Having the full shape information of four-point functions can thus help breaking degeneracy between different bias parameters. 
Last but not least, the computation of the trispectrum is required in order to accurately capture the covariance of the power spectrum, which encodes the statistical error information. An accurate account of the covariance will be important for realizing the full promise of the next generation LSS surveys.

\vskip 4pt
The impact of the non-Gaussian contribution to the power spectrum covariance from the connected four-point function has been studied in Fourier space e.g.~in~\cite{Scoccimarro:1999kp, Bertolini:2015fya, Bertolini:2016bmt, Mohammed:2016sre, Barreira:2017sqa, Barreira:2017kxd, Sugiyama:2019ike, Wadekar:2019rdu}. 
In contrast, no full computation of the covariance for the angular power spectrum has yet been performed due to the aforementioned numerical difficulties. 
In~\cite{Lacasa:2017ufk, Lacasa:2018hqp, Lazeyras:2017hxw}, the tree-level terms contributing to the covariance for galaxy clustering were derived and a subset of them were computed in the context of the halo model~\cite{Cooray:2002dia, Takahashi:2012em}. 
For weak lensing, the non-Gaussian covariance was computed using the flat-sky or Limber approximations~\cite{Scoccimarro:1999kp, Takada:2008fn, Barreira:2017fjz}, which are valid for small-angle sky coverage. 
However, the full validity of these approximations have not been tested, due to the lack of a stable method for computing the full angular four-point function without relying on these approximations.

\vskip 4pt
In this work, we present a method that bypasses these numerical difficulties by generalizing the previous FFTLog-based methods to angular four-point functions. In doing so, we revisit the separability condition of cosmological trispectra. Our method has a number of applications. First, it allows us to compute angular four-point functions with Gaussian and non-Gaussian initial conditions, the latter of which will be relevant for constraining primordial physics. Similarly, this can help constraining the cubic bias parameters that enters the galaxy four-point function at tree level. In addition, this FFTLog-based method provides a fast and reliable way of computing the non-Gaussian component of the covariance. We point out that, for the computation of the non-Gaussian covariance, the Limber approximation loses its validity even for high multipoles, in contrast to general expectations. We demonstrate this by computing the tree-level contribution to the covariance in standard cosmological perturbation theory.

\paragraph{Outline} The paper is organized as follows. In Section~\ref{sec:ang}, we describe an efficient method to compute cosmological angular four-point functions. In Section~\ref{sec:shapes}, we describe different trispectrum shapes that we consider in our analyses. We apply the method to compute the angular galaxy trispectrum in Section~\ref{sec:galaxy} and the non-Gaussian covariance of the angular matter power spectrum in Section~\ref{sec:cov}. We conclude in Section~\ref{sec:con}. A number of appendices contain supplementary and technical details. 
In Appendix~\ref{app:div}, we describe the method of dealing with spurious divergences. In Appendix~\ref{app:cubic}, we give details of cubic bias operators and their momentum space representation.  Finally, in Appendix \ref{app:spin} we present useful properties of spin-weighted functions on a sphere, and use them to write contact separable trispectra.

\paragraph{Notations and convention} 
We use $k_i=|\k_i|$ to denote the magnitudes of spatial momenta. 
The Fourier convention is $\tilde\O(\k) = \int_{\mathbb{R}^3} \d^3x\, e^{-i\k\cdot\x}\,\O(\x)$. The matter power spectrum is computed with \texttt{CLASS} using the best-fit parameters of the $\Lambda$CDM model from Planck 2018~\cite{Aghanim:2018eyx}: $h=0.674$, $\Omega_bh^2=0.0224$, $\Omega_ch^2=0.120$, $\tau=0.054$, $A_s=2.10\times 10^{-9}$ at $k=0.05\,{\rm Mpc}^{-1}$, and $n_s=0.965$.

\section{Methodology}\label{sec:ang}

We are interested in computing correlation functions on a two-dimensional sphere, so it is natural to expand them in spherical harmonics. The advantage of this decomposition is that it trivializes the angular integrations for statistically isotropic observables. At the same time, the challenge is that the radial integrals are difficult to evaluate, consisting of highly-oscillatory spherical Bessel functions. Having efficient algorithms for evaluating these integrals will be important for analyzing large sets of observational datasets from upcoming experiments.

\vskip 4pt
In this section, we present an efficient method for evaluating angular correlation functions, mainly focusing on the four-point case. For a self-contained discussion, we first briefly review the basics of angular correlations in cosmology in~\S\ref{sec:angcorr}. We then discuss the separability of correlators in~\S\ref{sec:sep}, reviewing the familiar case of bispectra and introducing a classification of separable trispectra. Lastly, we introduce a method to compute the angular trispectrum based on the FFTLog algorithm in~\S\ref{sec:angtri}.

\subsection{Angular Correlators}\label{sec:angcorr}

Many cosmological observables are measured in redshift space. These include the CMB, as well as weak lensing shear and galaxy density fields from redshift surveys. Suppose that there is an object located at some position $\x$ and redshift $z$ with respect to an observer sitting at the origin. The actual observable seen in the sky depends on the entire trajectory the light has undertaken from the object to reach the observer at redshift $z=0$. A cosmological observable $\O$ at redshift $z$ is thus defined as an integration along the line-of-sight direction $\hat\n \equiv \x/|\x|$ over some kernel $W_\O$ as
\begin{align}
	\O(\hat\n,z) = \int_0^\infty \d\chi\, W_\O(\chi)\O(\chi\hat\n,z) = \sum_{\ell m}\O_{\ell m}(z)Y_{\ell m}(\hat\n)\, ,\label{Olm}
\end{align}
where $\chi$ stands for comoving distance, and we have expanded the observable in spherical harmonics $Y_{\ell m}$, with $\sum_{\ell m}\equiv \sum_{\ell=0}^\infty\sum_{m=-\ell}^\ell$. The kernel can be either a sharp or broad function depending on the observable under consideration. 

\begin{figure}[t]
    \centering
         \includegraphics[scale=.9]{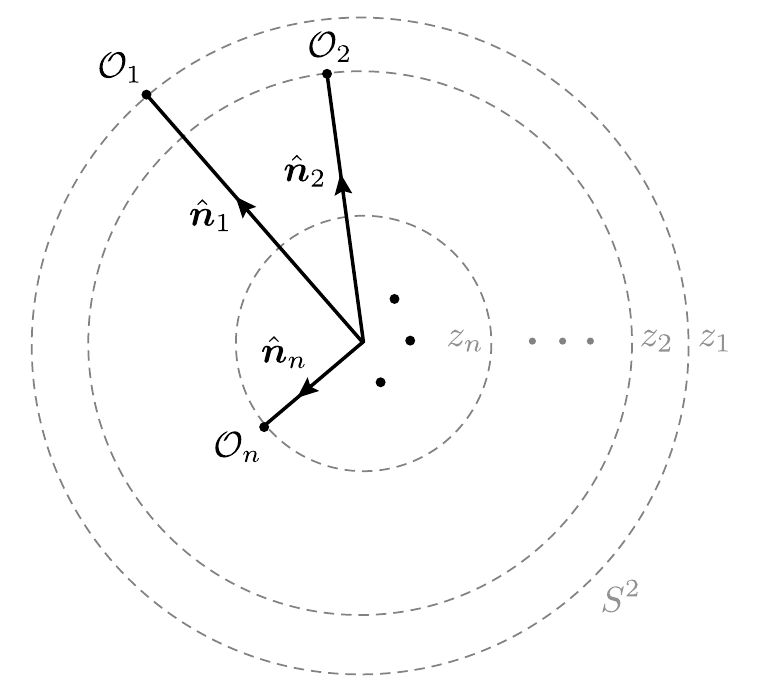}
         \\
    \caption{Illustration of the angular $n$-point function of observables $\O_i$ on celestial two-spheres $S^2$ in real space as a function of redshifts $z_i$ and line-of-sight directions $\hat\n_i$.}
    \label{fig:angcorr}
\end{figure}

\vskip 4pt
At fixed redshift, it is natural to characterize the statistics of the observable $\O$ in terms of the angular variable $\O_{\ell m}$ on the sphere.  
Cosmological observables are, however, usually first computed in Fourier space. Taking the Fourier transform of $\O$ and using the spherical harmonics expansion of the plane wave\footnote{By $Y_{\ell m}(\hat \k)$, we mean the spherical harmonic as a function of the angles of $\hat\k$ with respect to some fixed reference frame, which becomes irrelevant once we integrate over the angles.}
\begin{equation}
	e^{i\k\cdot\r} = 4\pi \sum_{\ell m} i^\ell j_\ell(kr)Y_{\ell m}^*(\hat\k)Y_{\ell m}(\hat\r)\, ,\label{eq:rayleigh}
\end{equation}
where $j_\ell$ is the spherical Bessel function, the projected observable can be expressed as
\begin{align}
	\O_{\ell m}(z) = 4\pi i^\ell \int_0^\infty \d\chi\, W_\O(\chi)\int_{\mathbb{R}^3}\frac{\d^3k}{(2\pi)^3}\, j_\ell (k\chi) Y_{\ell m}^*(\hat\k)\tilde\O(\k,z)\, ,\label{Olm2}
\end{align}
where $\tilde\O$ denotes the Fourier conjugate of the observable $\O$. The corresponding angular $n$-point function is then (see Fig.~\ref{fig:angcorr})
\begin{equation}
	\langle \O_{\ell_1 m_1}\cdots \O_{\ell_n m_n}\rangle = (4\pi)^ni^{\ell_{1\cdots n}}\int\!\left[\prod_{i=1}^n\frac{\d\chi_i\d^3k_i}{(2\pi)^3}\, W_\O(\chi_i)j_{\ell_i}(k_i\chi_i)Y_{\ell_im_i}^*(\hat\k_i)\right]\!\langle \tilde\O_1\cdots \tilde\O_n\rangle\, ,\label{angnpt}
\end{equation}
where we have suppressed the redshift dependence on the left-hand side. We defined $\ell_{1\cdots n}\equiv \ell_1+\cdots +\ell_n$, and 
\begin{align}
	\langle \tilde\O_1\cdots \tilde\O_n\rangle = \langle \tilde\O_1\cdots \tilde\O_n\rangle' \times (2\pi)^3\Ddelta(\k_1+\cdots+\k_n)\, \label{Ocorr}
\end{align}
denotes an $n$-point function in Fourier space, with $\tilde\O_i\equiv \tilde\O_i(\k_i,z_i)$, $\delta_{\rm D}$ is the Dirac delta function that enforces momentum conservation, and $\langle\cdots\rangle'$ is a correlator with the delta function stripped off. The above formula~\eqref{angnpt} tells us how to take a correlator in $k$-space and project it onto $\ell$-space.

\vskip 4pt
In this work, we consider angular correlators of galaxy density field $\delta_g$ at different redshifts, although the method can be similarly applied to other angular observables. These are related to matter density field $\delta$ by means of a bias expansion, which we describe in \S\ref{sec:bias}. There are two sources of matter density fields: (i) late-time gravitational nonlinearities and (ii) non-Gaussian initial conditions, with the latter being characterized by correlators of the primordial curvature perturbation $\zeta$. The former contribution can be computed using the framework of standard perturbation theory, which we briefly describe in \S\ref{subsec:grav}. The relation between $\delta$ and $\zeta$ is given by
\begin{align}
	\delta(\k,z) = \M(k,z)\zeta(\k)\, ,\quad \M(k,z) = -\frac{2}{5}\frac{k^2T(k)D_g(z)}{\Omega_{m,0} H_0^2}\, ,\label{transfer}
\end{align}
where $\M$ is the transfer function that evolves $\zeta$ to $\delta$ at redshift $z$, with normalization $T(0)=1$, and $D_g$ is the linear growth factor normalized to unity at $z=0$, ${D_g(0)=1}$. 

\vskip 4pt
We see from \eqref{angnpt} that the computation of an $n$-point angular correlator in general requires evaluating $4n$ coupled integrals, which becomes highly intractable as $n$ increases. This is further exacerbated due to the presence of the spherical Bessel functions, which makes the radial integrands highly oscillatory and therefore difficult to numerically integrate. 
Performing the angular integrations involving spherical harmonics, however, can always be done straightforwardly, which leads to a form restricted by rotational invariance. For the power spectrum and the bispectrum, statistical isotropy implies that they can be written as
\begin{align}
	\langle\O_{\ell m}\O_{\ell' m'}\rangle &=  \delta_{\ell\ell'}\delta_{mm'}\hs C_{\ell}\, ,\label{eq:powerspectrum}\\[3pt]
	\langle\O_{\ell_1 m_1}\O_{\ell_2 m_2}\O_{\ell_3 m_3}\rangle  &= \G^{\ell_1\ell_2\ell_3}_{m_1m_2m_3} \hs b_{\ell_1\ell_2\ell_3}\, ,\label{eq:bispectrum}
\end{align}
where the geometric factor $\G^{\ell_1\ell_2\ell_3}_{m_1m_2m_3}$ called the Gaunt coefficient is defined as
\begin{align}
\G^{\ell_1\ell_2\ell_3}_{m_1m_2m_3} &\equiv \int_{S^2} \d\Omega_{\hat\n}\, Y_{\ell_1 m_1}(\hat\n)Y_{\ell_1 m_2}(\hat\n)Y_{\ell_1 m_3}(\hat\n)=g^{\ell_1\ell_2\ell_3}\begin{pmatrix}
		\ell_1 & \ell_2 &\ell_3 \\ m_1 & m_2 & m_3
	\end{pmatrix}\, ,\label{gaunt}\\
	g^{\ell_1\ell_2\ell_3}&\equiv \sqrt{\frac{(2\ell_1+1)(2\ell_2+1)(2\ell_3+1)}{4\pi}}\begin{pmatrix}
		\ell_1 & \ell_2 &\ell_3 \\ 0&0&0
	\end{pmatrix}\, ,\label{gfactor}
\end{align}
and round-bracket matrices denote Wigner 3-$j$ symbols. The physical degrees of freedom of the angular bispectrum are thus characterized by the {\it reduced bispectrum} $b_{\ell_1\ell_2\ell_3}$, which is a function of three multipoles.  

\vskip 4pt
Unlike for $n=2$ and $3$, rotational invariance does not uniquely fix the form of angular $n$-point functions for $n\ge 4$. This is simply due to the fact that there are multiple ways of choosing diagonal multipoles for higher-point functions. 
For instance, the angular trispectrum can be expressed in a rotationally invariant form as~\cite{Hu:2001fa}
\begin{align}
	\langle \O_{\ell_1 m_1}\cdots \O_{\ell_4 m_4}\rangle &=\sum_{LM}(-1)^{M} \begin{pmatrix}
		\ell_1 & \ell_2 &L \\ m_1 & m_2 & M
	\end{pmatrix}\begin{pmatrix}
		\ell_3 & \ell_4 & L \\ m_3 & m_4 & -M
	\end{pmatrix} T^{\ell_1\ell_2}_{\ell_3\ell_4}(L)\label{angtri} \\[5pt]
	&=\sum_{LM} (-1)^{M} \begin{pmatrix}
		\ell_1 & \ell_2 &L \\ m_1 & m_2 & M
	\end{pmatrix}\begin{pmatrix}
		\ell_3 & \ell_4 & L \\ m_3 & m_4 & -M
	\end{pmatrix} P^{\ell_1\ell_2}_{\ell_3\ell_4}(L)\, + (2\leftrightarrow 3)+(2\leftrightarrow 4) \, ,\nonumber
\end{align}
where in the second line we have decomposed the trispectrum into a sum over different channels.\footnote{For implications of statistical isotropy on general angular $n$-point functions, see~\cite{Mitsou:2019ocs}.} Since we are interested in the connected part of the four-point function, we will assume that the disconnected (Gaussian) part has been subtracted off. Note that $P^{\ell_1\ell_2}_{\ell_3\ell_4}(L)$ is not invariant under $\ell_1\leftrightarrow\ell_2$ or $\ell_3\leftrightarrow\ell_4$, but instead satisfies $P^{\ell_1\ell_2}_{\ell_3\ell_4}(L) = (-1)^{\ell_{12}+L}P^{\ell_2\ell_1}_{\ell_3\ell_4}(L)=(-1)^{\ell_{34}+L}P^{\ell_1\ell_2}_{\ell_4\ell_3}(L)$ due to the properties of the 3-$j$ symbols. To exhaust all the permutation symmetry, it can be broken apart into four permutations as
\begin{align}
	P^{\ell_1\ell_2}_{\ell_3\ell_4}(L) = t^{\ell_1\ell_2}_{\ell_3\ell_4}(L) +(-1)^{\ell_{1234}}t^{\ell_2\ell_1}_{\ell_4\ell_3}(L)+(-1)^{\ell_{34}+L}t^{\ell_1\ell_2}_{\ell_4\ell_3}(L)+(-1)^{\ell_{12}+L}t^{\ell_2\ell_1}_{\ell_3\ell_4}(L)\, ,\label{Pdecomp}
\end{align}
where $t^{\ell_1\ell_2}_{\ell_3\ell_4}(L)$ is called the {\it reduced trispectrum}. This is invariant under the exchange of the upper and lower indices, $t^{\ell_1\ell_2}_{\ell_3\ell_4}(L) = t^{\ell_3\ell_4}_{\ell_1\ell_2}(L)$. Later on, we will see that the geometric factor $g^{\ell_1\ell_2\ell_3}$ repeatedly shows up in the calculation, as in the bispectrum case. It is therefore convenient to further decomposed the reduced trispectrum as
\begin{align}
	t^{\ell_1\ell_2}_{\ell_3\ell_4}(L) = g^{\ell_1\ell_2 L}g^{\ell_1\ell_2 L}\Big(\tau^{\ell_1\ell_2}_{\ell_3\ell_4}(L) + \tau_{\ell_1\ell_2}^{\ell_3\ell_4}(L)\Big)\, .\label{SRtri}
\end{align}
We will call $\tau^{\ell_1\ell_2}_{\ell_3\ell_4}(L)$ the {\it super-reduced trispectrum}.\footnote{In \cite{Regan:2010cn}, the terminology ``extra-reduced trispectrum" was used to refer to $\tau^{\ell_1\ell_2}_{\ell_3\ell_4}(L) + \tau_{\ell_1\ell_2}^{\ell_3\ell_4}(L)$.} The full trispectrum can be built out of $4!=24$ permutations of this basic building block.

\begin{figure}[t]
    \centering
         \includegraphics[scale=.9]{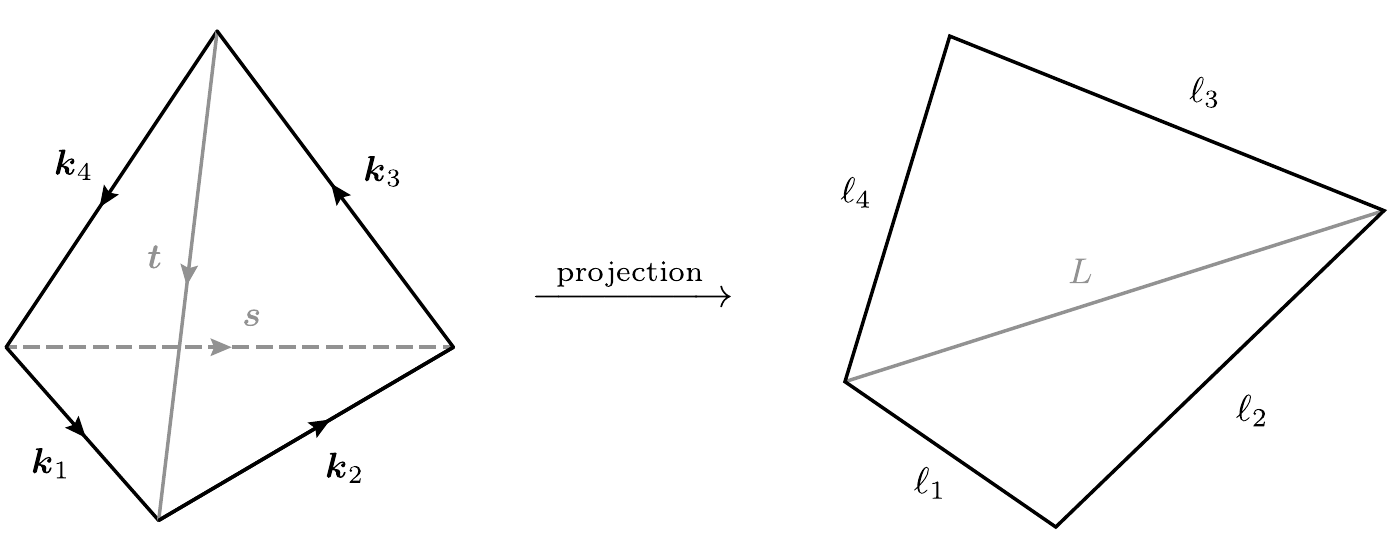}
         \\
    \caption{Kinematic configurations of four-point functions in $k$-space ({\it left}) and $\ell$-space ({\it right}). The internal momenta are denoted by $\s\equiv\k_1+\k_2$ and $\t\equiv\k_2+\k_3$, while the internal multipole for the pairing $\{\ell_1\ell_2,\ell_3\ell_4\}$ is denoted by $L$.}
    \label{fig:angcorr2}
\end{figure}

\vskip 4pt
Using various identities of the Wigner symbols, we can express the trispectrum in terms of a single pairing $\{\ell_1\ell_2,\ell_3\ell_4\}$ as
\begin{align}
	T^{\ell_1\ell_2}_{\ell_3\ell_4}(L) &= P^{\ell_1\ell_2}_{\ell_3\ell_4}(L)\\
	&\hskip -30pt  + (2L+1)\sum_{L'}  \Bigg( (-1)^{\ell_2+\ell_3}\begin{Bmatrix} \ell_1 & \ell_2 & L' \\ \ell_4 & \ell_3 & L \end{Bmatrix}P^{\ell_1\ell_3}_{\ell_2\ell_4}(L') +(-1)^{\ell_L+\ell_{L'}}\begin{Bmatrix} \ell_1 & \ell_2 & L' \\ \ell_3 & \ell_4 & L \end{Bmatrix}P^{\ell_1\ell_4}_{\ell_3\ell_2}(L')\Bigg) \, ,\nonumber
\end{align}
where the curly-bracketed matrices denote the Wigner 6-$j$ symbols. We see that the reduced trispectrum has five independent degrees of freedom, as opposed to six for the 4-point function in Fourier space. More generally, the total number of independent degrees of freedom for $n$-point functions $3n-6$ (for $n\ge 3$) in $k$-space gets reduced to $2n-3$ in $\ell$-space, which geometrically is described by an $n$-gon (see Fig.~\ref{fig:angcorr2}).

\subsection{Separability}\label{sec:sep}
A brute-force evaluation of \eqref{angnpt} for a general momentum-space correlator is highly intractable due to the sheer number of coupled multi-dimensional integrals. However, these integrals become greatly simplified for certain correlators that are {\it separable}, i.e.~for those that can be expressed as a product of functions of momenta. Let us first briefly review the separability of the bispectrum in \S\ref{sec:sepbi}, and then discuss an analogous criterion for the trispectrum in \S\ref{sec:septri}.

\subsubsection{Bispectrum: Review} \label{sec:sepbi}

Due to spatial isotropy, bispectra must be functions of the dot products of momenta $\k_i\cdot\k_j$, which can be traded with wavenumbers $k_i=|\k_i|$ by momentum conservation. 
A bispectrum in momentum space is then said to be separable if it can be expressed as
\begin{align}
	\langle\O(\k_1,z_1)\O(\k_2,z_2)\O(\k_3,z_3)\rangle' &=f_1(k_1,z_1)f_2(k_2,z_2)f_3(k_3,z_3)\, .\label{sep3pt}
\end{align}
In general, a full bispectrum will consists of a finite sum over such separable terms as well as other permutations. 

\vskip 4pt
It turns out that many physical bispectra can be expressed in the above separable form. 
To see why this is the case, let us first classify the possible shapes of primordial bispectra, for which there is no redshift dependence. Suppose first that the bispectrum is a rational function of momenta, so that it can be expressed as
\begin{align}
	\langle\zeta_{\k_1}\zeta_{\k_2}\zeta_{\k_3}\rangle' = \frac{G(k_1,k_2,k_3)}{H(k_1,k_2,k_3)}\, ,\label{PQ}
\end{align}
where $G$, $H$ are polynomials. The question of separability then depends on the form of $H$, which is determined in terms of its zeros or the singular behavior of the bispectrum, which can be either of the type $k_i\to 0$ or $k_1+k_2+k_3\to 0$.\footnote{This is true when the Bunch-Davies initial condition is imposed. Excited initial conditions can lead to singularities of the type $k_1+k_2-k_3\to 0$ that blow up in the folded configuration.} 
The former arises because the bispectrum is proportional to the power spectrum, while the latter is due to the kind of time integral involved in computing late-time correlators in inflation. This factor alone can be expressed as
\begin{align}
	\frac{1}{(k_1+k_2+k_3)^n} = \frac{1}{\Gamma(n)}\int_0^\infty \d\tau\, \tau^{n-1}e^{-(k_1+k_2+k_3) \tau}\, ,\label{kt}
\end{align}
which is simply the wick-rotated version of the time integral, which is numerically easier to evaluate than the Lorentzian version. 
A naively non-separable bispectrum containing a factor $(k_1+k_2+k_3)^{-n}$ can then be made separable by approximating the above integral by a finite sum~\cite{Smith:2006ud}.

\vskip 4pt
To see what the condition \eqref{sep3pt} implies for the angular bispectrum, we first expand the delta function in plane waves using 
\begin{equation}
	(2\pi)^3\Ddelta(\k_1+\cdots+\k_n) = \int_{\mathbb{R}^3}\d^3r \, e^{i\k_1\cdot\r}\cdots e^{i\k_n\cdot\r}\, ,\label{eq:deltaexpansion}
\end{equation}
and then project onto the spherical harmonics basis using \eqref{eq:rayleigh}. We can then easily perform the angular part of the integrals in \eqref{angnpt}. 
After stripping off the geometric factor in \eqref{angnpt}, the reduced bispectrum can then be expressed as
\begin{equation}
	b_{\ell_1\ell_2\ell_3} = \frac{1}{(2\pi^2)^3}\int_0^\infty\d r\, r^2 I_{\ell_1}^{(1)}(r)I_{\ell_2}^{(2)}(r)I_{\ell_3}^{(3)}(r)\, ,\label{reducedbis}
\end{equation}
where we defined
\begin{eBox}
\vskip 3pt
\begin{equation}
	I^{(i)}_{\ell}(r) \equiv 4\pi \int_0^\infty\d\chi\, W_\O(\chi) \int_0^\infty \d k\, k^{2}f_i(k,z(\chi)) j_\ell(kr)j_\ell(k\chi) \, .\label{Ifunc}
\end{equation}
\vskip 3pt
\end{eBox}
The expensive part of this calculation is the $k$-integral involving highly-oscillating spherical Bessel functions. Once this is done, however, the remaining $r$-integral has in general a smooth integrand, which can be replaced by a finite quadrature.

\subsubsection{Trispectrum}\label{sec:septri}
Having reviewed the well-known separable properties of bispectra, let us now discuss similar separability conditions for trispectra. (See \cite{Regan:2010cn, Smith:2015uia} for earlier works on separable trispectra.)
Analogous to the bispectrum case, one would be tempted to think that the condition
\begin{align}
	\langle \O(\k_1,z_1)\cdots\O(\k_4,z_4)\rangle' = f_1(k_1,z_1)\cdots f_4(k_4,z_4)\, ,
\end{align}
is a suitable definition of the separability for the trispectrum. This turns out to be too restrictive in general. To see why, let us briefly review the basic kinematics of trispectra. Spatial isometries imply that the number of independent degrees of freedom for an $n$-point function is $3n-6$ (for $n\ge 3$), or six for the trispectrum. 
It is natural to choose four of them to be the magnitudes of the external momenta, $k_i$ for $i=1,\cdots,4$. For our purposes, we will find it convenient to parameterize the remaining two degrees of freedom with the magnitudes of two of the three internal momenta, which we denote by the Mandelstam-like variables $s\equiv |\k_1+\k_2|$, $t\equiv |\k_2+\k_3|$, and $u\equiv |\k_1+\k_3|$. We introduce these variables by labelling the internal momenta with $\s$, $\t$, and $\u$ and then imposing the momentum conservation at each vertex. It is useful to first decompose the trispectrum into a sum over different channels as (temporarily dropping the $z$ dependence to avoid clutter)
\begin{align}
	\langle \O(\k_1,z_1)\cdots\O(\k_4,z_4)\rangle = P_\O(\k_1,\k_2,\k_3,\k_4)+(2\leftrightarrow 3)+(2\leftrightarrow 4)\, ,
\end{align}
where each channel can further be decomposed into
\begin{align}
	P_\O(\k_1,\k_2,\k_3,\k_4) = \tau_\O(\k_1,\k_2,\k_3,\k_4) +(1\leftrightarrow 2)+(3\leftrightarrow 4)+(12\leftrightarrow 34)\, ,
\end{align}
in analogy to \eqref{Pdecomp}. The trispectrum in the $s$-channel can then be written as
\beq	
\hskip -7pt\tau_\O(\k_1,\k_2,\k_3,\k_4)
	=\int\d^3s\int\d^3t \,\tau_\O(k_1,k_2,k_3,k_4,s,t)\hs\Ddelta(\k_{12}-\s)\Ddelta(\k_{23}-\t)\Ddelta(\k_{1234})\, ,\label{zeta4st}
\eeq
where $\k_{i_1\cdots i_n}\equiv \k_{i_1}+\cdots+\k_{i_n}$. In analogy to the bispectrum case, one can say that the trispectrum is separable if it can be written as
\be
\langle \O(\k_1,z_1)\cdots\O(\k_4,z_4)\rangle'= f_1(k_1,z_1)\cdots f_4(k_4,z_4)f(s)g(t)\, ,\label{generaltri}
\ee
where we have restored the redshift dependence, and $\langle\cdots\rangle'$ now indicates that the integrals over $\s$ and $\t$ with the delta functions are also stripped off. 
The separability condition for other permutations is obtained by a cyclic shift: $s\to t$, $t\to u$. 
As we explain in \S\ref{subsec:ng}, this choice of variables naturally parameterizes the shapes that arise from exchanging a mediator particle. For instance, the intermediate particle in the $s$-channel carries the momentum $\k_1+\k_2$ and the trispectrum becomes a polynomial in $t^2$ whose degree reflects the spin of the intermediate particle~\cite{Arkani-Hamed:2018kmz}. The spin-$J$ factor $t^{2J}$ can be replaced by the dot product between momenta in the cross channel, $(\hat \k_2\cdot \hat \k_3)^J$, or equivalently by a linear combination of the Legendre polynomials $P_J(\hat \k_2\cdot \hat \k_3)$. Combining these facts, we consider the following {\it ansatz} for a separable trispectrum: 
\begin{eBox}
\be
\langle \O(\k_1,z_1)\cdots\O(\k_4,z_4)\rangle'= f_1(k_1,z_1)\cdots f_4(k_4,z_4)f(s)\hs t^{2J}\, , \label{generaltri2}
\ee
\vskip 5pt
\end{eBox}
for the $s$-channel, where $J\ge 0$ is a non-negative integer. For the primordial trispectrum, the redshift dependence can be ignored.

\vskip 4pt
A general separable trispectrum will be a sum over separable pieces obeying~\eqref{generaltri2}. Given a separable trispectrum in Fourier space, we would like to see what form of the angular trispectrum that this leads to. As previously, the general strategy to derive angular correlators is as follows. We first use \eqref{eq:deltaexpansion} to express the delta functions as integrals over plane waves, which are then projected onto the spherical harmonic basis using \eqref{eq:rayleigh}. We then substitute the Fourier-space trispectrum to the projection formula \eqref{angnpt} and perform the integrations over the angles. This only leaves radial integrals to be evaluated, weighted by a geometric factor, which can be simplified and put in the standard isotropic form \eqref{angtri} using various identities involving the Wigner symbols. 

\vskip 4pt 
Let us now classify different types of separable trispectra, corresponding to different choices of $f(s)$ and $J$. We consider three cases illustrated by the following diagrams: 

\begin{figure}[!h]
\centering
\captionsetup[subfigure]{labelformat=empty, font=normalsize}
        \begin{subfigure}[b]{0.32\textwidth}
        \centering
                \includegraphics[height=2.8cm]{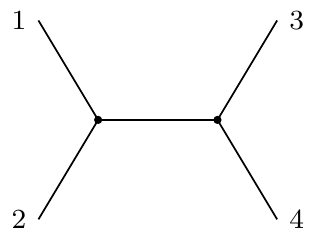}
                \caption{Scalar-Exchange (scE)\\[5pt] $f(s)$}
                \label{fig:scE}
        \end{subfigure}
        \begin{subfigure}[b]{.32\textwidth}
        \centering
                \includegraphics[height=2.8cm]{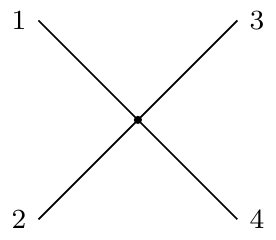}
                \caption{Contact (C)\\[5pt] $s^{2J}$}
                \label{fig:C}
        \end{subfigure}
        \begin{subfigure}[b]{.32\textwidth}
        \centering
                \includegraphics[height=2.8cm]{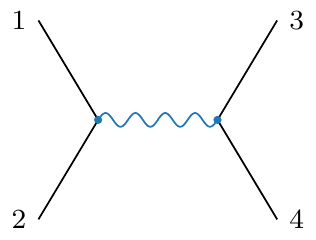}
                \caption{Spin-Exchange (spE)\\[5pt] $f(s)\hs t^{2J}$}
                \label{fig:spE}
        \end{subfigure}
\end{figure}
\vskip -4pt
\noindent
We have chosen suggestive names that reflect the physical processes that give rise to each shape dependence, as shown above. For example, scalar-exchange diagrams can have nontrivial dependence on the internal momentum, but with $J=0$. Contact diagrams can be viewed as a special case of the scalar-exchange diagram, where the dependence on the internal momentum is given by non-negative integer powers of $s^2$.  
Lastly, the spin-exchange diagrams generalize the scalar-exchange case to nonzero $J$.\footnote{A similar classification was used in~\cite{Smith:2015uia}, where scalar-exchange and spin-exchange separability were collectively referred to as ``exchange separability''. As we will see shortly, these two cases are qualitatively different, so it will be useful to present formulas for these two cases separately.}

\vskip 4pt
In what follows, we present formulas for the angular trispectra belonging to different separability classes in the $s$-channel. Answers for different channels can be obtained by permutations. Derivations involve straightforward algebra but are rather unilluminating, so we mostly just quote the results.

\paragraph{Scalar-Exchange (scE).}
First, we consider the case with $J=0$ and a generic function $f(s)$. Although contact separability leads to a simpler structure, we show the result for this case first because the contact-separable trispectrum can be derived as a special case of scE-separable one. We can rewrite the trispectrum \eqref{zeta4st} as
\begin{align}
	\langle \O_1\O_2\O_3\O_4\rangle =(2\pi)^3 f_1(k_1,z_1)\cdots f_4(k_4,z_4)\int_{\mathbb{R}^3} \d^3s\, f(s)\hs \Ddelta(\k_{12}-\s)\hs\Ddelta(\k_{34}+\s)\, ,\label{eq:case2}
\end{align}
where we have absorbed the total-momentum-conserving delta function $\Ddelta(\k_{1234})$ inside the $\s$-integral.\footnote{It is also possible not to absorb $\Ddelta(\k_{1234})$ in the $\s$-integral. However, after projection this leads to integrals involving three spherical Bessel functions, which are trickier to deal with.} 
Using the plane-wave expansion of the delta function, and then evaluating the angular integrals, 
and substituting into the projection formula~\eqref{angnpt}, we obtain the super-reduced trispectrum
\begin{eBox}
\vskip -2pt
\begin{equation}
	\hskip -8pt\text{(scE)}:\ \tau^{\ell_1\ell_2}_{\ell_3\ell_4}(L) = \frac{1}{(2\pi^2)^5}\!\int_0^\infty\! \d r\hs r^2I^{(1)}_{\ell_1}(r)I^{(2)}_{\ell_2}(r) \int_0^\infty\!\d r' r'^2   I^{(3)}_{\ell_3}(r')I^{(4)}_{\ell_4}(r')J_L^{(s)}(r,r')\hs ,\label{planarangtri}
\end{equation}
\vskip 3pt
\end{eBox}
where the function $I_\ell^{(i)}$ was defined in \eqref{eq:Iellr}. We see that the $s$-dependence of the momentum-space trispectrum leads to two coupled radial integrals, with the coupling integral given by
\begin{eBox}
\vskip 3pt
\begin{equation}
	J_L^{(s)}(r,r') \equiv 4\pi\int_0^\infty\d k\hs k^2f(k) j_L(kr)j_L(kr')\, .\label{Jkernel}
\end{equation}
\vskip 3pt
\end{eBox}
This integral basically has the same structure as the Bessel integral inside $I_\ell^{(i)}$, and is thus difficult to numerically evaluate for generic $L$. In the next section, we will see how this integral can be trivialized with the help of the FFTLog transform. The double integral in \eqref{planarangtri}, having smooth integrands, can then be numerically evaluated with a finite quadrature. 

\paragraph{Contact (C).}
The above result was valid for any function $f(s)$. Contact separability corresponds to the special case of scalar-exchange separability, where $f(s)=s^{2n}$ with non-negative integer $n$, which is also equivalent to setting $f(s)=1$ and $J=n$ in the $t$-channel. There are multiple ways of dealing with this case. Here we present a method that utilizes~\eqref{planarangtri}, which we find to give the most economical representation. 

\vskip 4pt
When $n$ is a non-negative integer the coupling integral \eqref{Jkernel} becomes divergent, but we can treat it as a distribution. For example, when $n=0$, \eqref{Jkernel} simply becomes proportional to the delta function $2\pi^2\hs\Ddelta(r-r')/r^2$ due to the closure relation for spherical Bessel functions. To deal with the case $n>0$, note that $j_\ell$ satisfies a differential equation~\cite{Assassi:2017lea}
\begin{align}
	\D_\ell(r) j_\ell(sr) = s^2 j_\ell(sr)  \quad\text{with}\quad \D_\ell(r)\equiv -\partial_r^2 - \frac{2}{r}\partial_r + \frac{\ell(\ell+1)}{r^2}\, ,\label{Dell}
\end{align}
so that we can formally express the integral as
\begin{align}
	J_L^{(s)}(r,r') = \frac{2\pi^2}{r'^2}\big[\D_\ell(r)\big]^n\Ddelta(r-r')\, .
\end{align}
We can then integrate by parts to act the $\D_\ell$ operator on the $r$ integrand, after which we impose the delta function to collapse the two radial integrals to a single one. Doing so, and stripping off the geometric factor, we find the super-reduced trispectrum to be
\begin{eBox}
\vskip 5pt
\begin{equation}
	\text{(C)}:\ \tau^{\ell_1\ell_2}_{\ell_3\ell_4}(L) =	\frac{1}{(2\pi^2)^4}\int_0^\infty\d r\, \big[\tilde\D_\ell(r)\big]^n\Big[r^2 I_{\ell_1}^{(1)}(r)I_{\ell_2}^{(2)}(r)\Big]I_{\ell_3}^{(3)}(r)I_{\ell_4}^{(4)}(r)\, ,\label{trisep2}
\end{equation}
\vskip -5pt
\end{eBox}
where
\begin{align}
	\tilde\D_\ell(r)\equiv -\partial_r^2 + \frac{2}{r}\partial_r + \frac{\ell(\ell+1)-2}{r^2}\, .
\end{align}
The derivatives of $\tilde\D_\ell$ can be taken either numerically or analytically as described in Appendix~\ref{app:div}. Since taking many derivatives can lead to numerical instabilities, in practice the derivatives are better taken symmetrically on the 1,2- and 3,4-legs at the same time. Since \eqref{trisep2} consists of a single radial integral, it has the same degree of computational complexity as the reduced bispectrum in~\eqref{reducedbis}.

\vskip 4pt
As a remark, let us mention that there is an alternative way of computing the trispectrum of the contact type via the use of spin-weighted spherical harmonics. The idea is to write factors of $s^{2n}$ in terms of the dot product $\k_1\cdot\k_2$, and then replace these with the radial derivatives acting on plane waves, e.g. ${\k_1\cdot\k_2\, e^{i\k_1\cdot\r}e^{i\k_2\cdot\r} = -\partial_{r_i}e^{i\k_1\cdot\r} \partial_{r_i}e^{i\k_2\cdot\r}}$, which in turn raise the spins of the spherical harmonics after projection. We give details of this method in Appendix~\ref{app:spin} (see~\cite{Smith:2015uia} for the application of this method for~$n=1$). However, this method quickly becomes complicated and we find it to be not easily generalizable for $n>1$. So in practice, it is better to use the representation~\eqref{trisep2}.

\paragraph{Spin-Exchange (spE).}
In this case, we organize the momentum-conserving delta functions in the same way as~\eqref{eq:case2}, with an extra integral for $\t$. As mentioned before, we can always write $t^{2J}$ in terms of a linear combination of the Legendre polynomials $P_m(\hat\k_2\cdot\hat\k_3)$ with $m=0,\cdots,J$. Via the addition theorem, the degree-$J$ Legendre polynomial can be written as
\begin{align}
	P_{J}(\hat\k_2\cdot\hat\k_3) = \frac{4\pi}{2J+1}\sum_{m=-J}^{J} Y_{Jm}(\hat \k_2)Y_{Jm}^*(\hat \k_3)\, .\label{eq:addition}
\end{align}
The presence of the extra spherical harmonics leads to a more complicated geometric factors compared to the scE-separable case when we perform the angular integrations. After a bit of algebra, we find that the super-reduced trispectrum takes the form (see also~\cite{Bordin:2019tyb})
\begin{eBox}
\vskip -5pt
\begin{align}
	\text{(spE)}:\ \tau^{\ell_1\ell_2}_{\ell_3\ell_4}(L) &= \sum_{L'\ell_1'\ell_3'}\frac{h^{\ell_1\ell_2}_{\ell_3\ell_4}(L,J,L',\ell_1',\ell_3')}{(2\pi^2)^5}\! \nn
	& \quad\times \int_0^\infty \d r\hs r^2I^{(1)}_{\ell_1\ell_1'}(r)I^{(2)}_{\ell_2}(r)\int_0^\infty\d r'\hs r'^2   I^{(3)}_{\ell_3\ell_3'}(r')I^{(4)}_{\ell_4}(r')J_L^{(s)}(r,r')\, ,\label{case3}
\end{align}
\vskip 1pt
\end{eBox}
where the geometric factor is given by
\begin{align}
	h^{\ell_1\ell_2}_{\ell_3\ell_4}(L,J,L',\ell_1',\ell_3')&\equiv \frac{4\pi(-1)^{\ell_{1234}+\frac{1}{2}(\ell_{13}'+\ell_{13})+L+L'+J}(2L+1)}{2J+1}\nn
	&\quad\times \frac{g^{\ell_1'\ell_3'L'}g^{\ell_2\ell_4L'}g^{\ell_1\ell_1'J}g^{\ell_3\ell_3'J}}{g^{\ell_1\ell_2 L}g^{\ell_3\ell_4 L}}\begin{Bmatrix}
		\ell_1 & \ell_2 & L \\ L' & J & \ell_1'
	\end{Bmatrix}\begin{Bmatrix}
		\ell_3 & \ell_4 & L \\ L' & J & \ell_3'
	\end{Bmatrix},
\end{align}
and the curly brackets denote the Wigner 6-$j$ symbol. The $g$-factors in the denominator are due to the way we defined $\tau^{\ell_1\ell_2}_{\ell_3\ell_4}(L)$ in \eqref{SRtri}. In the above, the integral
\begin{equation}
	I^{(i)}_{\ell\ell'}(r) \equiv 4\pi \int_0^\infty\d\chi\, W_\O(\chi) \int_0^\infty \d k\, k^{2}f_i(k,z(\chi)) j_\ell(k\chi)j_{\ell'}(kr) \, ,
\end{equation}
slightly generalizes \eqref{Ifunc} by allowing the two multipoles of the spherical Bessel functions to be different. Naively, one might worry about the appearance of the three extra summations in \eqref{case3}. However, it turns out most terms vanish in the sums due to the triangle conditions imposed by the 6-$j$ symbols.  As a consequence, the computation is not much slower than the scalar-exchange case. The performance will be discussed in detail in~\S\ref{sec:perf}.

\vskip 4pt
Note that when $f(s)=s^{2n}$ with non-negative integer $n$, \eqref{case3} can be simplified in the same manner as the contact-separable case by writing $J_L^{(s)}$ in terms of the delta function and then collapsing one of the radial integrals. In this case, \eqref{case3} simplifies to
\begin{equation}
	\hskip -5pt\tau^{\ell_1\ell_2}_{\ell_3\ell_4}(L) = \sum_{L'\ell_1'\ell_3'}\frac{h^{\ell_1\ell_2}_{\ell_3\ell_4}(L,J,L',\ell_1',\ell_3')}{(2\pi^2)^4} \int_0^\infty \d r\hs [\tilde\D_L(r)]^n\Big[ r^2I^{(1)}_{\ell_1\ell_1'}(r)I^{(2)}_{\ell_2}(r)\Big]  I^{(3)}_{\ell_3\ell_3'}(r')I^{(4)}_{\ell_4}(r)\, .\label{case4}
\end{equation}
As we describe in Section~\ref{sec:shapes}, this type of trispectrum can arise from higher-derivative self-interactions of $\zeta$, which would be generated by integrating out spinning particles that couple to $\zeta$ during inflation.

\subsection{Bessel Integrals}\label{sec:angtri}
In the previous section, we saw that a separable trispectrum leads to an integral over a product of factorized momentum integrals. The most challenging part of the computation is evaluating these momentum integrals consisting of an highly-oscillatory integrand. 
In this section, we describe an efficient method to compute these Bessel integrals based on the FFTLog algorithm, originally introduced in~\cite{Hamilton:1999uv}, and further developed in~\cite{McEwen:2016fjn, Assassi:2017lea, Simonovic:2017mhp, Gebhardt:2017chz, Schoneberg:2018fis}. 

\subsubsection{FFTLog}\label{sec:FFT}

The FFTLog is defined to be a discrete Fourier transform with $N_\eta$ logarithmically-spaced sampling points in the $k$-interval $[k_{\rm min},k_{\rm max}]$. Effectively, this decomposes functions into a sum of complex power-laws. 
For a given function $f(k,z)$, its FFTLog decomposition is given by\footnote{Although this is strictly speaking an approximate relation for a finite sum, for simplicity we will use an equal sign.}
\begin{align}\label{eq:FFTLog}
	f(k,z) = \sum_{m=-N_\eta/2}^{N_\eta/2}c_m(z) k^{-b+i\eta_m}\quad\text{with} \quad \eta_m\equiv \frac{2\pi m}{\log(k_{\rm max}/k_{\rm min})}\, ,
\end{align}
and the coefficients $c_m$ are given by the inverse transform
\begin{align}
	c_m(z)=\frac{2-\delta_{|m|,N_\eta/2}}{2N_\eta}\sum_{n=0}^{N_\eta-1}f(k_n,z)k_n^{b}k_{\rm min}^{-i\eta_m}e^{-2\pi i mn/N_\eta}\, ,
\end{align}
where the Kronecker delta ensures correct weighting factor at the end points $m=\pm N_\eta/2$. The parameter $b\in\mathbb{R}$ is inserted in order to ensure convergence of the FFTLog; see e.g.~\cite{McEwen:2016fjn, Assassi:2017lea} for more details.

\begin{figure}[t!]
    \centering
         \includegraphics[width=10.0cm]{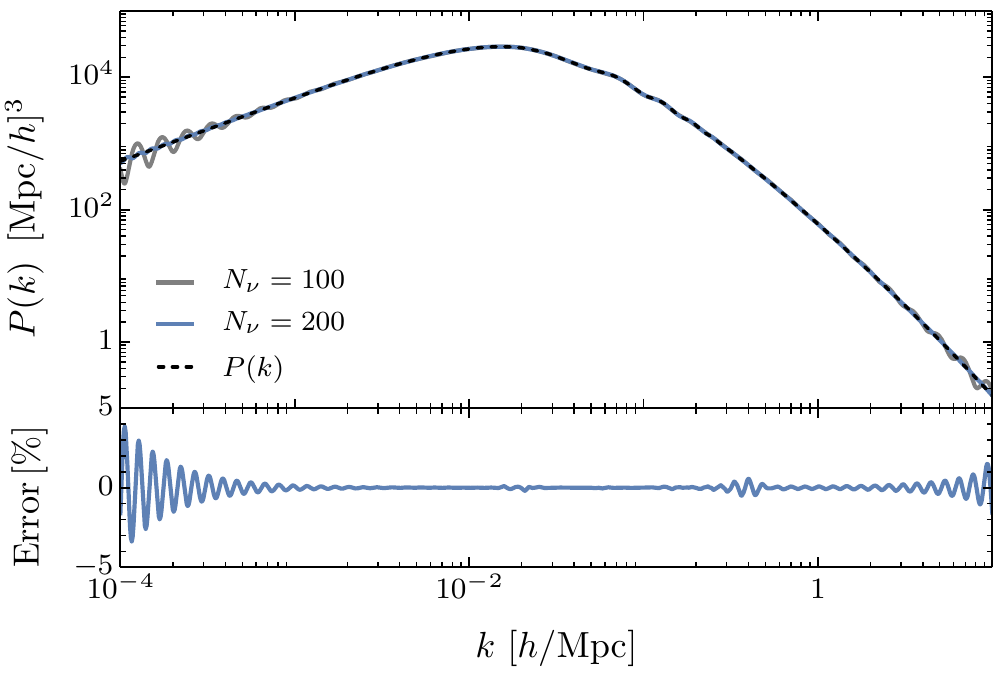}
         \\
    \caption{Comparison of the linear matter power spectrum and its FFTLog expansion with parameters $k_{\rm min}=10^{-4}$ $h/\text{Mpc}$, $k_{\rm max}=10^{2}$ $h/\text{Mpc}$, and $b=0.9$. The bottom panel shows the relative error for $N_\eta=200$ in percentage. }
    \label{fig:Pk}
\end{figure}

\begin{figure}[t]
    \centering
	\includegraphics[width=10.0cm]{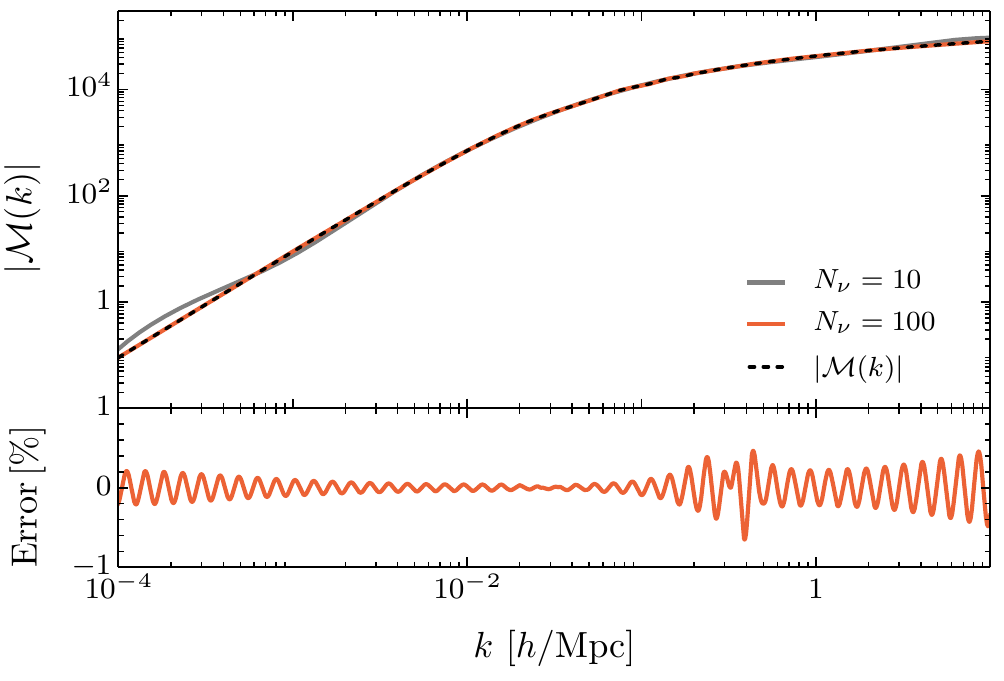}
         \\
    \caption{Comparison of the transfer function and its FFTLog expansion with parameters $k_{\rm min}=10^{-4}$ $h/\text{Mpc}$, $k_{\rm max}=10^{2}$ $h/\text{Mpc}$, and $b=-1.1$. The bottom panel shows the relative error for $N_\eta=100$ in percentage.}
    \label{fig:Mk}
\end{figure}

\vskip 4pt
In our analysis, we consider the FFTLog of two quantities: the matter power spectrum $P(k,z)$ and the transfer function $\M(k,z)$. These are relevant for computing angular correlators with Gaussian and non-Gaussian initial conditions, respectively. Let us show the FFTLog of these two quantities at redshift $z=0$ by taking $f(k,z)$ in \eqref{eq:FFTLog} to be $P(k)\equiv P(k,z=0)$ and $\M(k)\equiv \M(k,z=0)$, in which case the coefficients $c_m$ are $z$-independent. We use the \texttt{CLASS}\footnote{\url{https://class-code.net}} \cite{Lesgourgues:2011re} code to numerically compute these functions in the interval $k\in [10^{-5},10^2]$ $h/\text{Mpc}$.\footnote{Obviously, we cannot trust the linear approximation for the entire interval. We nevertheless choose a sufficiently large interval to avoid ringing in the FFTLog decomposition and to ensure convergence of momentum integrals. Since the integrals have support effectively on a finite interval of $k$ in the intermediate regime, they are not highly sensitive to the way the high- and low-$k$ limits are regulated; see also~\cite{Schoneberg:2018fis}.} Their asymptotic behaviors are given by
\begin{align}
	P(k)\, \propto\, \begin{cases}
		k^{n_s} & k\ll k_{\rm eq} \\ k^{n_s-4}\log^2 k \phantom{\hskip 20pt} & k\gg k_{\rm eq}
	\end{cases}\, ,
\end{align}
where $n_s\approx 0.96$ is the spectral index and $k_{\rm eq}\sim 10^{-2}\ {\rm Mpc}^{-1}$ is the scale corresponding to the matter-radiation equality. These behaviors set the allowed range of $b$: we find that the best convergence of the FFTLog is achieved when $b\in[0.5,2]$ and $b\in[-0.5,-1.5]$ for $P(k)$ and $\M(k)$, respectively. 
In Figs.~\ref{fig:Pk} and~\ref{fig:Mk}, we compare these functions and their FFTLog expansions. We see that we need about $N_\eta=200$ terms to require a percent-level precision for $P(k)$, whereas a less number $N_\eta=100$ is required for the convergence of $\M(k)$.

\subsubsection{Analytic Method}\label{sec:angres}

Let us see how the FFTLog can be used to efficiently compute angular galaxy correlators, following the method introduced in the earlier work~\cite{Assassi:2017lea}. We review their method in this section and extend to the cases involving the integral $J_L^{(s)}$ and primordial non-Gaussianity. 
Galaxies are measured over finite redshift bins, and due to errors in measuring their photometric redshifts, some of them may smear into other bins. This can be modeled with a Gaussian window function\footnote{For spectroscopic surveys, galaxy redshifts can be measured with much greater resolution, which makes a top-hat window function a more appropriate choice.}
\begin{align}
	W_\delta (\chi;\bar\chi,\sigma_\chi) \equiv \frac{1}{\sqrt{2\pi}\sigma_\chi}e^{-(\chi-\bar\chi)/2\sigma_\chi^2}\, ,\label{Wwindow}
\end{align}
where $\bar\chi$ is the mean redshift and $\sigma_\chi$ is the width of the bin. 
 When dealing with galaxy correlators, we will frequently encounter separable  coefficient functions of the form 
\begin{align}
	f_i(k,z) = k^{p_i}P(k)D_g(z)\, ,\quad f(k) = k^{p_s}P(k)\, ,
\end{align}
Substituting this to~\eqref{Ifunc} and \eqref{Jkernel} gives the integrals
\begin{align}
	I^{(i)}_{\ell}(r) &=4\pi \int_0^\infty\d\chi\, \W_\delta(\chi) \int_0^\infty \d k\, k^{2(1+p_i)}P(k) j_\ell(kr)j_\ell(k\chi)\, ,\label{Ifunc2}\\[5pt]
	J^{(s)}_{L}(r,r') &=4\pi \int_0^\infty \d k\, k^{2(1+p_s)}P(k) j_L(kr)j_L(kr')\, ,\label{Jkernel2}
\end{align}
where $\W_\delta(\chi)\equiv D_g(z(\chi))W_\delta(\chi;\bar\chi,\sigma_\chi)$. For high multipoles, these integrals can be trivially done using the Limber approximation~\cite{Limber:1954zz,LoVerde:2008re}. The idea is to note that the spherical Bessel function $j_\ell(x)$ becomes highly oscillatory for high $x>\ell$ and decays fast for small $x<\ell$, so that the integrand effectively becomes sharply peaked at $x\sim\ell$ for high $\ell$. This allows us to effectively replace the Bessel function as a Dirac delta function as
\begin{align}
	j_\ell(x)\to \sqrt{\frac{\pi}{2}}\frac{1}{\ell}\,\Ddelta(\ell-x)\quad \Rightarrow\quad  \begin{array}{ll}\displaystyle \hskip 15pt I_\ell^{(i)}(r) \to\frac{2\pi^2}{\ell^2}\W_\delta(r)\, , \\[10pt] \displaystyle  J_L^{(s)}(r,r') \to\frac{2\pi^2}{L^2}\, \left(\frac{L}{r}\right)^{2+2p_s}P(L/r)\, \Ddelta(r-r') \, . \end{array} \label{eq:limber}
\end{align}
This relies heavily on the assumption that the rest of the integrand is not highly varying over the integration region, so that the Delta function approximation is valid for high multipoles.

\vskip 4pt
Let us now see how the FFTLog can be used to evaluate the integral \eqref{Ifunc2} without the Limber approximation. Substituting the FFTLog of the matter power spectrum $P(k) = \sum_n c_n k^{-b+i\eta_n}$ to \eqref{Ifunc2} gives~\cite{Assassi:2017lea}
\begin{align}\label{eq:Iellr}
	I_\ell^{(i)}(r) = \sum_n c_n \int_0^\infty \d\chi\,   \W_\delta(\chi)\chi^{-\nu_n-2p_i} {\sf I}_\ell (\nu_n+2p_i,\tfrac{r}{\chi})\, ,
\end{align}
where we defined $\nu_n\equiv 3-b+i\eta_n$ and
\begin{eBox}
\vskip -5pt
\begin{align}
	{\sf I}_\ell(\nu,w) &\equiv 4\pi\int_0^\infty \d x\, x^{\nu-1}j_\ell(x)j_\ell(wx)\nn[3pt]
	&=\underbrace{\frac{2^{\nu-1}\pi^2\Gamma(\ell+\frac{\nu}{2})}{\Gamma(\frac{3-\nu}{2})\Gamma(\ell+\frac{3}{2})}\,w^\ell\, {}_2F_1\Bigg[\begin{array}{c} \frac{\nu-1}{2},\hs \ell+\frac{\nu}{2} \\[2pt] \ell+\frac{3}{2}\end{array}\Bigg|\, w^2\Bigg]}_{\equiv\, \tilde\I_\ell(\nu,w)}\quad (|w| \le 1)\, ,\label{WSint}
\end{align}
\vskip 0pt
\end{eBox}
with ${}_2F_1$ being the hypergeometric function. 
The second line $\tilde\I_\ell(\nu,w)$ represents the analytic solution for the integral valid when $|w|\le 1$, and the integral converges for  $-2\ell<\Re[\nu]<3$ when $w\ne 1$ and $-2\ell<\Re[\nu]<2$ when $w= 1$~\cite{Watson}. 
The function $\I_\ell(\nu,w)$ is discontinuous at $w=1$, invalidating a naive analytic continuation of the hypergeometric function appearing in \eqref{WSint} beyond the unit circle $|w|=1$. To compute the integral for $w>1$, we simply use the scaling property of the integral\footnote{
A naive analytic continuation of the hypergeometric function in \eqref{WSint} would lead to
\begin{equation}
	\tilde\I_\ell(\nu,w) = e^{-\frac{1}{2}i\nu\pi}w^{-\nu}\Big[\cos(\tfrac{\pi\nu}{2}) \I_\ell(\nu,\tfrac{1}{w}) - i\sin(\tfrac{\pi\nu}{2}) \I_{-\ell-1}(\nu,\tfrac{1}{w})\Big]\ \ (|w|\ge 1)\, ,
\end{equation}
which is the same as the right-hand side of \eqref{Iinverse} for even integer $\nu$ only. 
Other analytic properties of the hypergeometric function can still be used as long as we stay within the unit circle.
}
\begin{align}
	{\sf I}_\ell(\nu,w) = w^{-\nu}\hs {\sf I}_\ell(\nu,\tfrac{1}{w})\, ,\label{Iinverse}
\end{align}
The analytic representation \eqref{WSint} is very useful, since the hypergeometric series converges very fast; we refer the reader to~\cite{Assassi:2017lea, Gebhardt:2017chz, Schoneberg:2018fis} for more details on efficient evaluation of \eqref{WSint}. The fact that we can convert an integral over highly oscillating Bessel functions to a finite sum over hypergeometric series is what makes the FFTLog decomposition very powerful.

\vskip 4pt
When choosing $b\in [0.5,2]$ for the convergence of the FFTLog of $P(k)$, the spherical Bessel integral is be UV-divergent for $p_i>0$. 
One way of dealing with the non-convergent case is by treating the integral formally as a distribution and then repeatedly acting with a differential operator on a lower-order integral as~\cite{Assassi:2017lea}
\begin{align}
	\I_\ell(\nu+2n;w) &= \big[\D_{\ell}(w)\big]^n\hs \I_\ell(\nu,w)  \, ,\label{Iellderiv}
\end{align}
where $\D_\ell$ was defined in \eqref{Dell}. We can then integrate this operator by parts, after which the derivatives act on the window function without affecting the rest of the radial integral; in other words, we have
\begin{align}\label{eq:Iellr3}
	I_\ell^{(i)}(r) = \sum_n  c_n\int_0^\infty \d\chi\,   \big[\tilde\D_\ell^{p_i}(\chi)\W_\delta(\chi)\big]\chi^{-\nu_n}\hs {\sf I}_\ell (\nu_n,\tfrac{r}{\chi})\, ,
\end{align}
where we have dropped the boundary terms, which are negligible for the window function~\eqref{Wwindow}.\footnote{Note that the relation \eqref{Iellderiv} is in fact a valid identity of the hypergeometric function for any $\nu$, i.e.
\begin{align}
	\tilde\I_\ell(\nu+2n;w) &= \big[\D_{\ell}(w)\big]^n\hs \tilde\I_\ell(\nu,w)  \, ,\label{Iellderiv2}
\end{align}
which can be shown using the known identities for the hypergeometric function. Naively, this implies that integrating $\D_\ell$ by part is in principle not necessary, and we can simply use the left-hand side of \eqref{Iellderiv} to deal with the UV divergence. The problem is that the hypergeometric function has a singularity $\tilde\I_\ell(\nu,w)\to (1-w)^{2-\nu}$ as $w\to 1$ when $\Re[\nu]> 2$ (or a logarithmic singularity when $\Re[\nu]=2$), making the line-of-sight integral very sensitive near $\chi=r$. In general, we thus follow \eqref{eq:Iellr3}, although it turns out that using \eqref{Iellderiv2} can still approximately give the correct result when the singularity is somewhat mild, e.g.~for $\Re[\nu]\lesssim 5$.} Similarly, the coupling integral $J_L^{(s)}(r,r')$ for exchange-separable trispectra (c.f.~\eqref{planarangtri} and \eqref{case3}) can be expressed as
\begin{align}
	J_L^{(s)}(r,r') &= 4\pi\int_0^\infty\d k\hs k^{2+2p_s}P(k) j_L(kr)j_L(kr')\nn
	&=\big[\D_L(r)\big]^{p_s}\sum_n c_n  r'^{-\nu_n}\I_L(\nu_n,\tfrac{r}{r'})\, .
\end{align}
Notice that the operator $\D_L(r)$ here depends on $r$ instead of $\chi$. Integrating this operator by parts will then hit $I_{\ell_i}^{(i)}$, 
similar to \eqref{trisep2} but without producing a delta function. The numerical computation of the radial integrals then simply reduces to a matrix multiplication for a finite array of $J_L^{(s)}(r,r')$.

\vskip 4pt
In the case of primordial non-Gaussianity, we will deal with 
\begin{align}
	f_i(k,z) = k^{2p_i+\alpha_i}\M(k)D_g(z)\, ,
\end{align}
with $\alpha_i\in\{0,n_s-4\}$, which follows from the relation \eqref{transfer} and the $\zeta$ power spectrum, $P_\zeta(k) \propto k^{n_s-4}$. The integral~\eqref{Ifunc} then becomes
\begin{align}
	I^{(i)}_{\ell}(r) &=4\pi \int_0^\infty\d\chi\, \W_\delta(\chi) \int_0^\infty \d k\, k^{2(1+p_i)+\alpha_i}\M(k) j_\ell(kr)j_\ell(k\chi)\, .\label{Ifunc2M}
\end{align}
The FFTLog of the transfer function $\M(k) = \sum_n \tilde c_n k^{-\tilde b+i\tilde\eta_n}$ then gives
\begin{align}
	I_\ell^{(i)}(r) = \sum_n \tilde c_n \int_0^\infty \d\chi\,   \W_\delta^{(1)}(\chi)\chi^{-\tilde\nu_n-2p_i-\alpha_i} {\sf I}_\ell (\tilde\nu_n+2p_i+\alpha_i,\tfrac{r}{\chi})\, ,
\end{align}
where $\tilde\nu_n\equiv 3-\tilde b+i\tilde\eta_n$. For the convergence of the FFTLog, we choose $\tilde b\in [-0.5,-1.5]$. This implies that the integral is UV-divergent for $\alpha_i=0$, which can be dealt with following the same procedure outlined above.

\section{Shapes of Trispectra}\label{sec:shapes}

In this section, we present the expressions of the shapes of the trispectra, for both Gaussian and non-Gaussian initial conditions, and identify their separability types according to the classification introduced in the precious section. We first briefly summarize the gravitationally-induced trispectrum in~\S\ref{subsec:grav}. We then describe a few physically-motivated primordial trispectra in~\S\ref{subsec:ng}.

\subsection{Non-Gaussianity from Gravitational Evolution}\label{subsec:grav}

Gravitational attraction is a nonlinear process. Statistics of density perturbations in the late-universe thus become non-Gaussian even if they were initially Gaussian distributed. A well-established formalism to compute the cosmological evolution of density perturbations is standard perturbation theory (SPT), see~\cite{Bernardeau:2001qr} for a review. Here we present the bare minimum of SPT required for describing the gravitationally-induced trispectrum at tree level, i.e.~at leading order in perturbation theory.\footnote{It is well-known that SPT fails to be consistent beyond tree level. In order to continue to make sense of perturbation theory at loop level, other formalisms have been developed such as the effective field theory of large-scale structure~\cite{Baumann:2010tm, Carrasco:2012cv}. As our analysis is restricted to tree level, SPT will be sufficient for our purposes.}

\vskip 4pt
Treating dark matter as a pressureless fluid, the evolution of the matter density field $\delta$ and its velocity divergence $\theta\equiv\nabla\cdot\v$ is described by the continuity and the Euler equations.
For small $\delta$ and $\theta$, the equations of motion can be solved perturbatively as an expansion in the linear solution $\delta^{(1)}(\k,z) = D_g(z) \delta_0(\k)$, where $\delta_0$ is the density field at $z=0$. Similarly, we have $\theta^{(1)}(\k,z) = -\H f_g(z)D_g(z) \delta_0(\k)$ where $f_g\equiv \d \log D_g/\d \log a$ denotes the logarithmic derivative of $D_g$ with respect to the scale factor $a$. Under the approximation $f_g\approx\Omega_m^{1/2}$, the nonlinear solutions can be written as power series in $\delta_0$, separately for $\delta$ and $\theta$ as
\begin{align}
	\delta(\k,z) &= \sum_{n=1}^\infty D_g(z)^n\delta^{(n)}(\k,z)\, ,\\
	\theta(\k,z) &= -\H \Omega_m^{1/2}\sum_{n=1}^\infty D_g(z)^n\theta^{(n)}(z)\, ,
\end{align}
with the $n$-th order solutions given by
\begin{align}
	\delta^{(n)} (\k) &= \int_{\q_1,\cdots,\q_n}(2\pi)^3\delta_D(\k-\q_{1\cdots n}) F_n^{\rm sym}(\q_1,\cdots,\q_n) \delta_0(\q_1)\cdots \delta_0(\q_n)\, ,\\
	\theta^{(n)} (\k) &= \int_{\q_1,\cdots,\q_n}(2\pi)^3\delta_D(\k-\q_{1\cdots n})G_n^{\rm sym}(\q_1,\cdots,\q_n) \delta_0(\q_1)\cdots \delta_0(\q_n)\, ,
\end{align}
where $\int_{\q_1,\cdots,\q_n}\equiv(2\pi)^{-3n}\int\d^3q_1\cdots\d^3q_n$ and $F_n^{\rm sym}$, $G_n^{\rm sym}$ denote symmetrization of the kernels $F_n$, $G_n$ that can be computed iteratively using the formulas given in~\cite{Goroff:1986ep, Jain:1993jh}. For example, we have $F_1=G_1=1$ and
\begin{align}
	F_2^{\rm sym}(\k_1,\k_2) &=\frac{5}{7}+ \frac{\k_1\cdot\k_2}{2k_1k_2}\left(\frac{k_1}{k_2}+\frac{k_2}{k_1}\right)+\frac{2}{7}\frac{(\k_1\cdot\k_2)^2}{(k_1k_2)^2}\, , \label{F2sym}\\
	G_2^{\rm sym}(\k_1,\k_2) &=\frac{3}{7}+ \frac{\k_1\cdot\k_2}{2k_1k_2}\left(\frac{k_1}{k_2}+\frac{k_2}{k_1}\right)+\frac{4}{7}\frac{(\k_1\cdot\k_2)^2}{(k_1k_2)^2}\, .\label{G2sym}
\end{align}

\vskip 4pt
The computation of correlation functions of $\delta$ proceeds in an analogous way as computing Feynman diagrams. At tree-level, there are two contributions to the matter trispectrum assuming Gaussian initial conditions. For $\delta(\k)\equiv\delta(\k,z=0)$, two different contractions yield
\begin{align}
	\langle\delta(\k_1)\delta(\k_2)\delta(\k_3)\delta(\k_4)\rangle' = T_{2211}(\k_1,\k_2,\k_3,\k_4) + T_{3111}(\k_1,\k_2,\k_3,\k_4)\, ,\label{mattertri}
\end{align}
where
\begin{align}
	T_{2211}(\k_1,\k_2,\k_3,\k_4) &= 4F_2^{\rm sym}(\k_{12},-\k_2)F_2^{\rm sym}(\k_{12},\k_3)P(k_{12})P(k_2)P(k_3) + \text{11 perms}\, ,\label{T2211mat}\\[3pt]
	T_{3111}(\k_1,\k_2,\k_3,\k_4)  &= 6 F_3^{\rm sym}(\k_1,\k_2,\k_3)P(k_1)P(k_2)P(k_3)+\text{3 perms}\, ,\label{T3111mat}
\end{align}
and $F_3^{\rm sym}$ denotes the symmetrized version of the SPT kernel
\begin{align}
	F_3(\k_1,\k_2,\k_3) &= \frac{1}{126}\left(\frac{3\k_1\cdot\k_{12}}{k_1^2}+\frac{2(\k_1\cdot\k_2)k_{12}^2}{(k_1k_2)^2}\right)\left(\frac{7\k_{12}\cdot\k_{123}}{k_{12}^2}+\frac{(\k_{12}\cdot\k_3)k_{123}^2}{k_{12}^2 k_3^2}\right)\\[3pt]
	&\hskip -50pt +\frac{1}{18}\bigg[\frac{\k_1\cdot\k_{123}}{k_1^2}\left(\frac{5\k_2\cdot\k_{23}}{k_2^2}+\frac{(\k_2\cdot\k_3)k_{23}^2}{(k_2k_3)^2}\right)+\frac{(\k_1\cdot\k_{23})k_{123}^2}{7k_1^2k_{23}^2}\left(\frac{3\k_2\cdot\k_{23}}{k_2^2}+\frac{(2\k_2\cdot\k_3) k_{23}^2}{(k_2k_3)^2}\right)\bigg]\, , \nonumber
\end{align}
with $k_{i_1\cdots i_n}\equiv |\k_{i_1}+\cdots+\k_{i_n}|$. To see what separability class this trispectrum falls into, we first convert the dot products $\k_i\cdot\k_j$ to diagonal momenta. For $F_2^{\rm sym}(\k_{12},-\k_2)$ in \eqref{T2211mat}, we can express it as
\begin{align}
	28F_2^{\rm sym}(\k_{12},-\k_2) &= \left(3k_2^2-5k_1^2+\frac{2k_2^4}{k_1^2}\right)\frac{1}{s^2}+\left(10+\frac{3k_2^2}{k_1^2}\right)-\frac{5s^2}{k_1^2}\, .\label{F2symT2211}
\end{align}
Similarly, $F_2^{\rm sym}(\k_{12},\k_3)=F_2^{\rm sym}(-\k_{34},\k_3)$ also depends on $s$ but not $t$. Combined with $P(k_{12})=P(s)$ in \eqref{T2211mat}, we see that $T_{2211}$ is scalar-exchange separable. Now consider $T_{3111}$. Naively, $F_3(\k_1,\k_2,\k_3)$ contains products of different permutations of dot products, so that there could be terms that depend on two diagonal momenta at the same time. However, using momentum conservation one can show that each term in $T_{3111}$ depends no more than one diagonal momentum. We therefore see that the gravitationally-induced matter trispectrum is scalar-exchange separable.

\vskip 4pt
It turns out that the unsymmetrized kernel $F_3(\k_1,\k_2,\k_3)$ depends both on $s$ and $t$, even though there are no products between these factors, so that it is still scalar-exchange separable. For the purpose of computing (reduced) angular trispectra, it will be slightly more convenient to rearrange the $F_3$ kernel in a way that makes the symmetry between different channels more manifest. To this end, we define a related kernel $\hat F_3$, which when symmetrized gives the same result as $F_3$, i.e.~$F_3^{\rm sym}(\k_1,\k_2,\k_3) = \hat F_3^{\rm sym}(\k_1,\k_2,\k_3)$, but each permutation of which depends only on a single internal momentum. We give its precise definition in \eqref{F3hat}, and use this basis of kernel henceforth.

\subsection{Non-Gaussianity from Initial Conditions}\label{subsec:ng}
We now consider the shapes of a few primordial trispectra generated during inflation (see~\cite{Biagetti:2019bnp} for a recent review). When evolved to late times, the $\zeta$ trispectrum is related to the matter trispectrum by
\begin{align}
	\langle\delta(\k_1,z_1)\cdots \delta(\k_1,z_1)\rangle' = \M(k_1,z_1)\cdots \M(k_4,z_4)\langle\zeta(\k_1)\cdots \zeta(\k_4)\rangle'\, ,
\end{align}
at tree level. Since multiplying by transfer functions does not induce any diagonal momentum dependence, here we discuss separability types of primordial trispectra.

\paragraph*{Local shape.}
A simple parameterization of non-Gaussianity is given by a local expansion of Gaussian random fields in real space
\begin{align}
	\zeta(\x) = \zeta_{\rm G}(\x)+ \frac{3}{5}f_{\rm NL}(\zeta_{\rm G}^2(\x)-\langle\zeta_{\rm G}^2(\x)\rangle) + \frac{9}{25}g_{\rm NL}(\zeta_{\rm G}^3(\x)-\langle\zeta_{\rm G}^3(\x)\rangle)\, ,
\end{align}
with Gaussian $\zeta_{\rm G}$. These terms lead to two contributions to the local trispectrum given by
\begin{align}
	\langle\zeta^4\rangle'_{\tnl} &= \tnl \Big[P_\zeta(k_1)P_\zeta(k_3)P_\zeta(|\k_1+\k_2|) + \text{11 perms}\Big]\, ,\label{tnlT} \\[5pt]
	\langle\zeta^4\rangle'_{\gnl} &= \frac{54}{24}\, \gnl \Big[P_\zeta(k_1)P_\zeta(k_2)P_\zeta(k_3) + \text{3 perms}\Big]\, ,\label{gnlT}
\end{align}
where $\langle\zeta^4\rangle\equiv\langle\zeta(\k_1)\cdots \zeta(\k_4)\rangle$. The current observational bounds from the CMB\footnote{See~\cite{Munshi:2009wy, Munshi:2010bh, Regan:2010cn, Fergusson:2010gn} for the construction of trispectrum estimators.} are $\gnl=(-5.8\pm 6.5)\times 10^4$ (68\% C.L.)~\cite{Akrami:2019izv} and $\tnl< 2.8\times 10^3$ (95\% C.L.)~\cite{Ade:2013ydc}. 
From their momentum dependence, we see that the $\tnl$ trispectrum is scalar-exchange separable, while the $\gnl$ trispectrum is contact separable.

\vskip 4pt
In single-field models of inflation, a nonzero bispectrum in the squeezed limit (at which the local shape peaks) necessarily generates the trispectrum of the $\tau_{\rm NL}$ shape whose amplitude saturates\footnote{This holds for models in which the dominant contribution to non-Gaussianity is given by a single, non-gravitational source.} the Suyama-Yamaguchi bound $\tnl \ge (\tfrac{6}{5} \fnl)^2$~\cite{Suyama:2007bg}, while $g_{\rm NL}$ is an independent variable that characterizes quartic self-interactions of the inflaton field. In contrast, certain non-single-field models---such as multi-field inflation~\cite{Byrnes:2008zy}, quasi-single-field inflation~\cite{Assassi:2012zq} or models with higher-spin fields~\cite{Baumann:2017jvh}---can generate a large trispectrum with $\tnl \gg (\tfrac{6}{5} \fnl)^2$. In the large-scale structure, these models can be constrained by stochasticity in the bias expansion~\cite{Baumann:2012bc, Ferraro:2012bd} (see also~\cite{Desjacques:2016bnm, An:2017rwo}).

\paragraph*{Equilateral shape.}
Another category of primordial trispectra involves the shapes generated by quartic self-interactions in the inflationary action. At leading order in derivatives, the three quartic interactions that contribute to the trispectrum are $\dot\sigma^4$, $\dot\sigma^2(\partial_i\sigma)^2$, and $(\partial_i\sigma)^4$ in the effective field theory of inflation~\cite{Cheung:2007st}, where $\sigma$ is some additional light scalar~\cite{Senatore:2010wk}. These lead to the following shapes of the trispectrum~\cite{Smith:2015uia}:
\begin{align}
	\langle\zeta^4\rangle_{\rm eq,1}'&=\frac{221184}{25}\gnl^{\text{eq},1}\frac{1}{k_1k_2k_3k_4 (k_1+k_2+k_3+k_4)^5}\, ,\nonumber\\[5pt]
\langle\zeta^4\rangle_{\rm eq,2}'&=-\frac{27648}{325}\gnl^{\text{eq},2}\frac{k_t^2+3(k_3+k_4)k_t+12k_3k_4}{k_1k_2(k_3k_4)^3(k_1+k_2+k_3+k_4)^5}(\k_3\cdot\k_4)+\text{5 perms}\, ,\label{Teq123}\\[5pt]
\langle\zeta^4\rangle_{\rm eq,3}'&=\frac{165888}{2575}\gnl^{\text{eq},3}\frac{2k_t^4-2k_t^2\sum_i k_i^2+k_t\sum_i k_i^3+12k_1k_2k_3k_4}{(k_1k_2k_3k_4)^3(k_1+k_2+k_3+k_4)^5}\big((\k_1\cdot\k_2)(\k_3\cdot\k_4)+\text{2 perms}\big)\, ,\nonumber
\end{align}
whose amplitudes are normalized such that $\langle\zeta^4\rangle' = \frac{216}{25}\gnl P_\zeta(k)^3$ in the tetrahedral configuration where $k_i=k$ and $\hat\k_i\cdot\hat\k_j=-1/3$ for $i\ne j$. As before, the factors of $(k_1+k_2+k_3+k_4)^5$ in the denominator can be made separable using the trick \eqref{kt}, after which the shapes become contact separable.\footnote{In this case, we would have factors such as $f_i(k) = k^\nu e^{-k\alpha}$ appearing in the momentum integrals. It turns out that there exists an analytic formula for the Bessel integral in this case as well, which is given by~\cite{Bateman, Gradshteyn} 
\begin{align}
	&{\sf J}_{\ell\ell'}(\nu,\alpha,w) \equiv 4\pi\int_0^\infty \d x\, x^{\nu-1}e^{-\alpha x} j_\ell(x)j_{\ell'}(wx)\nn
	&=\frac{w^{\ell'}}{2^{\ell+\ell'}\ell'!\alpha^{\ell+\ell'+\nu}} \sum_{m=0}^\infty \frac{\Gamma(\ell+\ell'+\nu+2m)}{m!(\ell+m)!(-2\alpha)^{2m}}\, {}_2F_1\Bigg[\begin{array}{c} -\ell-m,\hs -m \\[2pt] 1+\ell'\end{array}\Bigg|\, w^2\Bigg]\quad (|w|\le 1,\, \alpha>1)\, .\label{Jfunc2}
\end{align}
For integer $\ell,\ell'$, the hypergeometric function above can be identified as the Jacobi polynomial, i.e.~the summand above is a finite polynomial of $w^2$. 
Unfortunately, this analytic formula is not valid for small $\alpha \lesssim 1/x$, from which the integral actually receives a dominant contribution. In practice, this means that the above formula should be used in conjunction with ordinary numerical integration within a small range. Nevertheless, the formula \eqref{Jfunc2} is still useful, since it allows us to circumvent the oscillatory part of the integral.} 
In single-field models, a large trispectrum may also be generated from higher-derivative interactions~\cite{Huang:2006eha, Chen:2009bc, Arroja:2009pd, Behbahani:2014upa, Bartolo:2010di, Creminelli:2010qf, Bartolo:2013eka, Arroja:2013dya}. Trispectra in these models contain higher powers of momentum dot products and generally also fall into the contact-separable type. 



\paragraph*{Exchange shape.}

When there are extra particles that couple to the inflaton during inflation, they can produce distinct shapes of non-Gaussianity that carry information about the masses and spins of the particles. This allows a model-independent way of constraining the particle spectrum during inflation, akin to collider searches for new particles in particle accelerators. The physics of these non-Gaussian correlators was highlighted in~\cite{Arkani-Hamed:2015bza} and the resulting phenomenology has been greatly explored, see e.g.~\cite{Chen:2009zp, Baumann:2011nk, Assassi:2012zq, Noumi:2012vr, Lee:2016vti, Chen:2016uwp, Chisari:2016xki, MoradinezhadDizgah:2017szk, Kehagias:2017cym, Kumar:2017ecc, An:2017hlx, An:2017rwo, Baumann:2017jvh, Chen:2018sce, MoradinezhadDizgah:2018ssw, Liu:2019fag, Alexander:2019vtb, Hook:2019vcn, MoradinezhadDizgah:2019xun}.

\vskip 4pt
An analytic expression for the exchange four-point function in inflation was presented in~\cite{Arkani-Hamed:2018kmz, Baumann:2019oyu}, assuming weak couplings to the inflaton. Schematically, it has the following structure in the $s$-channel:
\begin{equation}
\hskip-5pt	\langle\zeta^4\rangle'_{\rm ex} = \frac{g^2\, s^3}{(k_1k_2k_3k_4)^{3/2}}\sum_{m=0}^J \Pi^{(J,m)}(\alpha,\beta,\tau)U_{12}^{(J,m)}(\partial_u,u,\alpha)U_{34}^{(J,m)}(\partial_v,v,\beta) \hat F(u,v)\, ,\label{Tex}
\end{equation}
with the kinematic variables defined by
\begin{equation}
	u\equiv \frac{s}{k_1+k_2}\, ,\ \ v\equiv \frac{s}{k_3+k_4}\, , \ \ \alpha\equiv k_1-k_2\,,\ \ \beta\equiv k_3-k_4\, , \ \ \tau\equiv {(\k_1-\k_2)\cdot(\k_3-\k_4)}\, ,
\end{equation}
and a coupling constant $g$. In the above, $\Pi^{(J,m)}$ is a polarization structure, which depends on the angular variable $\tau$ as $\Pi^{(J,m)}\sim \tau^m$, and $U_{ij}^{(J,m)}$ are second-order differential operators that acts on the seed function $\hat F(u,v)$ that encodes the scalar-exchange shape. From the relation $\tau=k_1^2+k_2^2+k_3^2+ k_4^2-s^2-2t^2$, we see that the spin-$J$ exchange trispectrum in the $s$-channel is a degree-$J$ polynomial in $t^2$. 

\vskip 4pt
The exchange trispectrum~\eqref{Tex}, being a function of $u$ and $v$, is not manifestly separable. However, its functional form is dramatically simplified in certain kinematic configurations: In the collapsed limit $s\to 0$, the operators $U_{ij}^{(J,m)}$ become trivial, and the shape dependence reduces to
\begin{equation}\label{eq:Tex_spin}
	\langle\zeta^4\rangle'_{\rm ex}\, \xrightarrow{s\,\to\, 0} \, \frac{g^2}{(k_1k_3s)^3}\left(\frac{s^2}{k_1k_3}\right)^{\frac{3}{2}+ i\mu}\sum_{\lambda=0}^J c_{J,\lambda}(\mu) Y_{J\lambda}(\hat\k_1)Y_{J\lambda}^*(\hat\k_3)+c.c.\, ,
\end{equation}
where {\it c.c.} stands for {\it complex conjugate}. This has a clean physical interpretation as indicating particle production during inflation, where the parameter $\mu\sim M/H$ refers to the mass of the particle in Hubble units during inflation. The mass-dependent coefficient $c_{J,\lambda}(\mu)$ is fixed by conformal symmetry~\cite{Arkani-Hamed:2015bza, Arkani-Hamed:2018kmz}, which goes as $c_{J,\lambda}(\mu)\sim e^{-\pi\mu}$ for $\mu \gg 1$.\footnote{In the effective field theory of inflation context, these coefficients are fixed in terms of the propagation speeds of individual helicity modes~\cite{Bordin:2018pca}.} Away from the collapsed limit, the shape becomes dominated by the equilateral shapes in \eqref{Teq123} for large masses.
For data analysis purposes, it is then convenient to approximate the exchange trispectrum with the following template given by a sum over two contributions as
\begin{align}
	\langle\zeta^4\rangle'_{\rm ex}\, \approx \,\langle\zeta^4\rangle'_{\rm ex}\Big|_{s\to 0}+ r(\mu) \hs\langle\zeta^4\rangle'_{\rm eq}\, ,
\end{align}
where $T_{\rm eq}$ is a linear combination of the equilateral trispectra in \eqref{Teq123} with some relative coefficient $r(\mu)$ that goes as $r(\mu)\sim 1/\mu^2$ for $\mu\gg 1$. See~\cite{MoradinezhadDizgah:2018ssw} for a similar template constructed for the bispectrum. This is a good approximation in the large-mass regime, but breaks down for $\mu\lesssim 1$, in which case the collapsed-limit shape receives corrections from a tower of higher-derivative shapes. In this regime, one should instead use the full shape given by \eqref{Tex}.

\section{Angular Galaxy Trispectrum}\label{sec:galaxy}

In this section, we compute the angular galaxy trispectrum using the FFTLog-based method described in Section~\ref{sec:ang}. The computation of the angular trispectrum beyond the Limber approximation in this section is new to this paper and have not been computed before.\footnote{In addition to the galaxy clustering trispectrum, there are other types of angular trispectra in the large-scale structure that are also sourced by gravitational nonlinearities such as the lensing trispectrum, studied e.g.~in~\cite{Foreman:2018gnv, Schaan:2018yeh}. Combining the information from clustering and lensing correlation functions (as well as their cross-correlations) is important for extracting optimal  cosmological constraints.} We briefly review the cubic bias expansion of galaxy density fields in \S\ref{sec:bias}, and describe how to implement redshift space distortion (RSD) in \S\ref{sec:RSD}. We present the shapes of the angular galaxy trispectrum with and without non-Gaussian initial conditions in \S\ref{sec:res}.
	
\subsection{Cubic Bias}\label{sec:bias}
We do not observe the distribution of matter density directly, but rather that of its tracers such as galaxies. 
One way of relating the galaxy density field $\delta_g$ to that of matter is to express the former as a local functional of operators that characterize the underlying matter density. 
The $n$-th order galaxy density field is then given by an expansion in a set of operators 
\begin{align}
	\delta_g^{(n)}(\x,z) =\sum_{\O\hs\in\hs\O_n} b_{\O}\hs \O^{(n)}(\x,z)\, ,\label{eq:bias}
\end{align}
where the superscript of $\O^{(n)}$ indicates that it is $n$-th order in the linear density $\delta^{(1)}$, $\O_n$ denotes a set of independent $n$-th order operators, and the coefficients $b_\O$ are called {\it bias parameters} or simply {\it biases}, which are redshift-dependent in general. 

\vskip 4pt
The operators that appear in the bias expansion can be classified  by the number of fields and their derivatives. A set of independent operators that appear at cubic order is~\cite{Chan:2012jj, Assassi:2014fva} 
\begin{align}
	{\cal O}_3 = \big\{ \delta,\,\delta^2,\,\G_2(\Phi_g),\,\delta^3,\, \G_2(\Phi_g)\delta,\, \G_3(\Phi_g),\,\Gamma_3 \big\}\, ,\label{O3}
\end{align}
where $\Gamma_3\equiv \G_2(\Phi_g)-\G_2(\Phi_v)$ and $\G_i$ are the Galileon operators defined by
\begin{align}
	\G_2(\Phi_g) &\equiv (\nabla_i\nabla_j\Phi_g)^2-(\nabla^2\Phi_g)^2\, ,\\[3pt]
	\G_3(\Phi_g) &\equiv \frac{3}{2}(\nabla_i\nabla_j\Phi_g)^2\nabla^2\Phi_g-(\nabla_i\nabla_j\Phi_g)(\nabla_j\nabla_k\Phi_g)(\nabla_k\nabla_i\Phi_g)-\frac{1}{2}(\nabla^2\Phi_g)^3\, ,
\end{align}
with $\Phi_g\equiv \nabla^{-2}\delta$ and $\Phi_v\equiv \nabla^{-2}\theta$ being the gravitational and velocity potentials, respectively. Of course, the choice of this set is not unique, and one could equally choose a set given by a linear combinations of the operators in \eqref{O3}. The relation to some other sets of operators that are also used in the literature is described in Appendix~\ref{app:cubic}.

\vskip 4pt
Using the bias expansion, we find that the tree-level galaxy trispectrum with Gaussian initial conditions has the form
\begin{align}
	\langle \delta_g(\k_1,z_1)\cdots\delta_g(\k_4,z_4)\rangle' &= T_{2211}^g(\k_1,\k_2,\k_3,\k_4)+T_{3111}^g(\k_1,\k_2,\k_3,\k_4)\, ,
\end{align}
with two contributions given by
\begin{align}
	T_{2211}^g(\k_1,\cdots,\k_4) &= 4b_\delta^2D_1D_2^2D_3D_4^2P(k_1)P(k_3)P(s) \Big(b_\delta F_2^{\rm sym}(\k_1,-\k_{12})+b_{\delta^2}+b_{\G_2}\sigma_{\k_1,-\k_{12}}^2\Big) \nn[3pt]
	& \times \Big(b_\delta F_2^{\rm sym}(\k_3,\k_{12})+b_{\delta^2}+b_{\G_2}\sigma_{\k_3,\k_{12}}^2\Big) + \text{11 perms}\, ,\label{T2211g}
	\end{align}
	\begin{align}
	T_{3111}^g(\k_1,\cdots,\k_4)&= b_\delta^3D_1D_2D_3D_4^3\, P(k_1)P(k_2)P(k_3)\nn[3pt]
	&\times \bigg\{ 6\Big[b_\delta \hat F_3^{\rm sym}(\k_1,\k_2,\k_3)+b_{\delta^3}-b_{\G_3} (\hat\k_1\cdot\hat\k_2)(\hat\k_2\cdot\hat\k_3)(\hat\k_3\cdot\hat\k_1)\Big] \nn[3pt]
	&+\Big[(b_{\G_2\delta}+\tfrac{3}{2}b_{\G_3})\sigma^2_{\k_1,\k_2}  + 2(b_{\G_2}+b_{\Gamma_3})\sigma^2_{\k_{12},\k_3}F_2^{\rm sym}(\k_1,\k_2),\label{T3111g}\\[3pt]
	&+ 2b_{\delta^2}F_2^{\rm sym}(\k_1,\k_2)-2b_{\Gamma_3}\sigma^2_{\k_{12},\k_3}G_2^{\rm sym}(\k_1,\k_2) + \text{5 perms}\Big] \bigg\} + \text{3 perms}\, ,\nonumber
\end{align}
where $D_i\equiv D_g(z_i)$ and $\sigma^2_{\k,\q} \equiv (\hat\k\cdot\hat\q)^2-1$. Let us classify the terms that appear in the trispectrum according to the types of separability we introduced in \S\ref{sec:septri}. First of all, we see that terms that arise from $T_{2211}^g$ all have a nontrivial dependence on the internal momentum, $s=|\k_{12}|$ for the particular permutation shown above. In particular, $T_{2211}^g$ does not depend on $t$, implying that it is scalar-exchange separable. The only spin-exchange separable term is that due to $b_{\G_3}$ in $T_{3111}^g$ that depends on two independent angles, with the rest being either contact or scalar-exchange separable depending on their momentum dependence. 
We summarize the separability classes of the trispectra from different bias parameters (as well as that of primordial non-Gaussianity) in Table~\ref{table:class}.

\vskip 5pt
\begin{table}[h!]
\begin{center}
\begin{tabular}{cccc}
\tline\\[-10pt]
\hskip 10pt Trispectrum \phantom{\hskip 5pt} & \phantom{\hskip 5pt}Scalar-Exchange\phantom{\hskip 5pt}  &\phantom{\hskip 5pt} Contact \phantom{\hskip 5pt}&\phantom{\hskip 5pt} Spin-Exchange\phantom{\hskip 10pt} \ \\[3pt]
\tline\rowcolor[gray]{0.92}{} &  & & \\[-10pt]
	\rowcolor[gray]{0.92}{} Biases & $b_{\delta}$, $b_{\delta^2}$, $b_{\G_2}$, $b_{\Gamma_3}$ & $b_{\delta^3}$, $b_{\G_2\delta}$ & $b_{\G_3}$ \\[5pt]
	\rowcolor[gray]{0.92}{} &  & & \\[-10pt]
	\rowcolor[gray]{0.92}{} Primordial NG & $\tnl^{\rm loc}$, $\tnl^{\rm scalar}$ & $\gnl^{\rm loc}$, $\gnl^{\rm eq}$ & $\tnl^{\rm spin}$ \\[5pt]
\tline
\end{tabular}
\end{center}
\vspace{-0.5cm}
	\caption{Separability classes of trispectra from cubic biases and primordial non-Gaussianity.}
	\label{table:class}
\end{table}

\subsection{Redshift Space Distortion}\label{sec:RSD}

What we actually measure in galaxy surveys are the fluctuations $\Delta(\hat\n,z)$ in galaxy number counts $N(\hat\n,z)$ defined by
\begin{align}
	\Delta(\hat\n,z)\equiv \frac{N(\hat\n,z)-\langle N(\hat\n,z)\rangle}{\langle N(\hat\n,z)\rangle}\, .
\end{align}
This is a gauge-invariant quantity that should encapsulate all relativistic effects. At leading order in perturbation theory, this is related to the density fluctuation in Fourier space as
\begin{align}
	\Delta^{(1)}(\hat\k,z) = \delta^{(1)}(\hat\k,z)\Big[b_\delta+f_g(z)(\hat\k\cdot\hat\n)^2\Big]\, .\label{RSD}
\end{align}
This correction is the standard redshift-space distortion term that arises from peculiar velocities of galaxies, known as the Kaiser effect~\cite{Kaiser:1987qv}. The full inclusion of all relativistic effects can be found in~\cite{Yoo:2009au, Yoo:2010ni, Bonvin:2011bg, Challinor:2011bk}, but here we consider a simple prescription by just keeping this RSD term. The correction in \eqref{RSD} leads to a change in the $I_\ell^{(i)}$ integral \eqref{Ifunc} as~\cite{Assassi:2017lea}
\begin{align}
	I_\ell^{(i)}(r) &\to 4\pi\int_0^\infty \d\chi\, \W_g(\chi)\int_0^\infty\d k\,k^2 \Big[b_\delta j_\ell(k\chi)-f_g(\chi)j_\ell''(k\chi)\Big]j_\ell(kr) f_i(k)\nn[3pt]
	&= 4\pi\int_0^\infty \d\chi\, \left[b_\delta \tilde\D_\ell(\chi) \W_g(\chi)-(f_g\cdot \W_g)''(\chi)\right]\int_0^\infty\d k\, j_\ell(k\chi)j_\ell(kr) f_i(k)\, ,
\end{align}
where in the second line we have integrated by parts to trade the derivatives of the spherical Bessel function with that of the window function, and $(f_g\cdot \W_g)''(\chi)\equiv \frac{\d^2}{\d \chi^2}(f_g(\chi)\W_g(\chi))$.

\begin{figure}[t!]
    \centering
         \includegraphics[width=\textwidth]{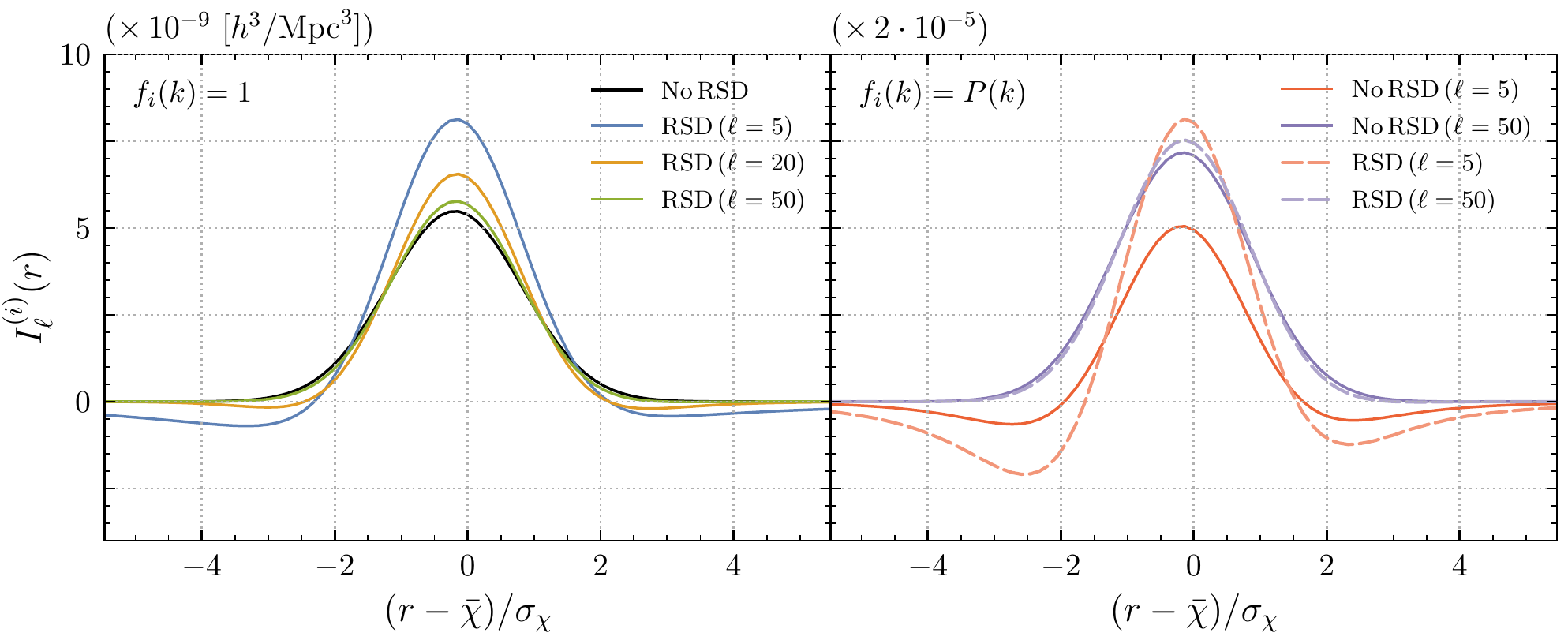}
         \\
    \caption{Comparison of $I_\ell^{(i)}$ with and without redshift space distortion (RSD) for $f_i(k)=1$ ({\it left}) and $f_i(k)=P(k)$ ({\it right}). The window function is centered at $z=1$ with $\sigma_z=0.1$.}
    \label{fig:RSDint}
\end{figure}

\vskip 4pt
When $f_i(k)=k^{2p_i}$, the integral simplifies. Using the identity
\begin{align}
	4\pi\int_0^\infty \d k\, j_\ell(k\chi)j_\ell(kr) = \frac{\pi^2}{\ell+\frac{1}{2}}\frac{1}{r}\left(\frac{\chi}{r}\right)^\ell \quad (\chi<r)\, ,
\end{align}
we have
\begin{align}
	\hskip -7pt I_\ell^{(i)}(r)&\to \begin{cases}\displaystyle \frac{2\pi^2}{r^2}\tilde\D_\ell^{p_i-1}(r)\Big[b_\delta\tilde\D_\ell(r)\W_g(r) - (f_g\cdot \W_g)''(r)\Big] & p_i>0\\[15pt]
\displaystyle \frac{2\pi^2}{r^2}b_\delta\W_g(r)\hs{-}\hs\frac{\pi^2}{\ell+\frac{1}{2}}\int_0^1\! \d x\hs x^\ell\bigg[(f_g\cdot \W_g)''(rx)   +  \frac{(f_g\cdot \W_g)''(r/x)}{x^3} \bigg] & p_i=0 \end{cases} .\label{IintRSD}
\end{align}
In Fig.~\ref{fig:RSDint}, we show the integral $I_\ell^{(i)}$ with and without RSD for $f_i(k)=1$ and $f_i(k)=P(k)$. Notice that the two different integrals have different units, but these cancel out in the end so that the final angular correlator is dimensionless. For $f_i(k)=1$, the integral without RSD simply collapses into a window function and is therefore independent of $\ell$. We see that the RSD effect becomes suppressed at high multipoles, as can also be seen from \eqref{IintRSD}. When $f_i(k)=P(k)$, the overall shape of the integral still looks like a window function. Again, the RSD effect shrinks for high multipoles.

\subsection{Shapes of Angular Trispectra}\label{sec:res}

We now consider the shape of the galaxy trispectrum in angular space. To better illustrate the shape contribution of individual terms, we compute the super-reduced trispectrum in the $s$-channel. For example, the contribution of the bias parameter $b_\delta$ in \eqref{T3111g} to the reduced trispectrum is 
\begin{align}
	\hskip -15pt\tau_{2211}^{\delta}(\k_1,\k_2,\k_3,\k_4) &= 2b_\delta^4 D_1 D_2^2D_3D_4 P(k_1)P(k_3)P(s) \big[F_2^{\rm sym}(\k_1,-\k_{12})F_2^{\rm sym}(\k_3,\k_{12})\big],\\[3pt]
	\tau_{3111}^{\delta}(\k_1,\k_2,\k_3,\k_4) &= \frac{1}{6}\hs b_\delta^4 D_1 D_2D_3D_4^3 P(k_1)P(k_2)P(k_3)\big[\hat  F_3(\k_1,\k_2,\k_3)\big]\hs .\label{That3111}
\end{align}
From \eqref{F2symT2211}, we see that $t_{2211}$ is a linear combination of terms
\begin{align}
	\tau_{2211}^{\delta}(\k_1,\k_2,\k_3,\k_4)\ \supset\ \big[D_1k_1^{2p_1}P(k_1)\big]\big[D_2k_2^{2p_2}P(k_2)\big]\big[D_3^2k_3^{2p_3}\big]\big[D_4^2k_4^{2p_4}\big]\big[s^{2p_s}P(s)\big]\, ,\label{T2211}
\end{align}
with $p_1,p_2\in\{-1,0,1\}$, $p_3,p_4\in\{0,1,2\}$, and $p_s\in\{-2,-1,0,1,2\}$. Note that we have terms that go as $s^{-4}P(s)$, so that the integral over $s$ is naively IR-divergent for $L=0$. This is because the integrand goes as $s^2 j_L(rs)j_L(r's)[s^{-4}P(s)]\sim  s^{-b-2+2L}$ as $s\to 0$; for $b\in[0.5,2.0]$ that is required for the convergence of the FFTLog decomposition, the integral diverges for $L=0$. However, this divergence is spurious, since we have not done the radial integrals that impose the momentum conservation between $s$ and other momenta. This suggests that these divergences could be removed by shifting factors of $k_i^2$ and $s^2$ amongst different integrals that would render the final integral finite, similar to the way we used to deal with the UV divergences in \S\ref{sec:angres}. We verify in Appendix~\ref{app:div} that this IR divergence can indeed be removed in this way.

\vskip 4pt
The expression \eqref{F3hat} of $F_3^{\rm sym}$ implies that $\tau_{3111}^{b_\delta}$ consists of terms as
\begin{align}
	\tau_{3111}^{\delta}(\k_1,\k_2,\k_3,\k_4)\ \supset\ \big[D_1k_1^{2p_1}P(k_1)\big]\big[D_2k_2^{2p_2}P(k_2)\big]\big[D_3k_3^{2p_3}P(k_3)\big]\big[D_4^3k_4^{2p_4}\big]\big[s^{2p_s}\big] \, .
\end{align} 
Again, the momentum integrals in this case become divergent for certain values of $p_i$, which can be dealt with in the same way as the above case. Note that the dependence on the diagonal momentum simply involves integer powers of $s$. In the special case $p_s=0$, the $s$ integral just becomes delta function, and the double radial integral collapses to a single integral. When $p_s=-1$, we can use the analytic formula for the two Bessel integral
\begin{align}
	4\pi\int_0^\infty \d s\, j_L(rs)j_L(r's) = \frac{\pi^2}{\ell+\frac{1}{2}}\frac{1}{r'}\left(\frac{r}{r'}\right)^L\quad (r<r')\, .
\end{align}
For $p_s> 1$, we lower $p_s$ by the use of the operator $\D_\ell$. It is straightforward to read off the super-reduced trispectra for other bias parameters, which can be analyzed in the same way as above. A complete list of all of them can be found in Appendix~\ref{app:cubic}.

\begin{figure}[t!]
    \centering      
      \includegraphics[width=\textwidth]{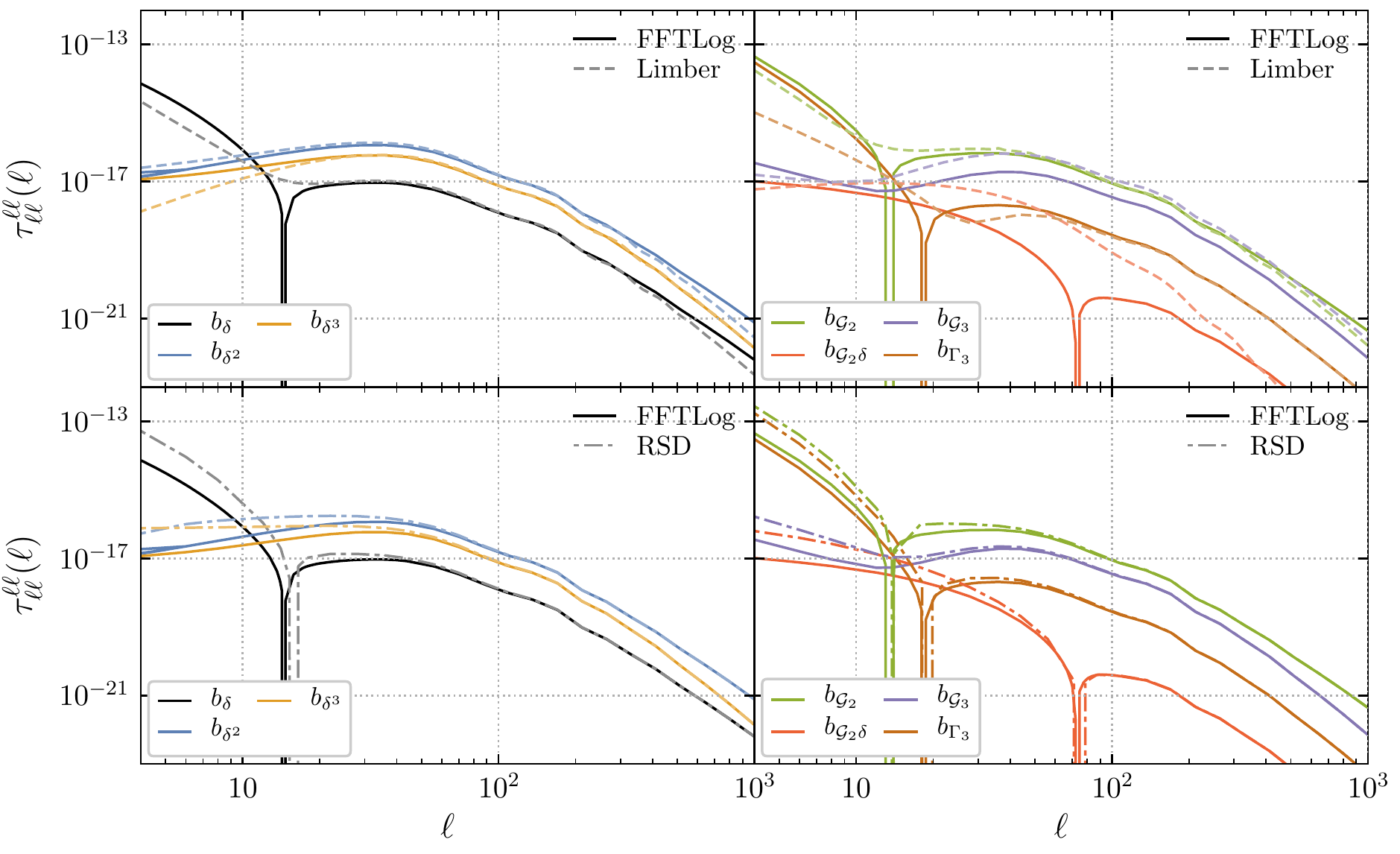}\\                
    \caption{Angular galaxy trispectrum in the equilateral configuration, $\ell_i=L=\ell$. The trispectrum is evaluated with the window function centered at redshift $z=1$ with $\sigma_z=0.1$. We used $N_\eta=200$ terms in the FFTLog expansion of $P(k)$ with $b=1.9$, and $N_\chi=50$, $N_r=100$ sampling points to numerically evaluate the integrals. In the upper panel, we compare the angular trispectrum computed using the FFTLog method (solid line) and the Limber approximation (dashed line). In the lower panel, we compare the angular trispectrum with (dot-dashed line) and without RSD (solid line), both computed using the FFTLog method. Different colors indicate the trispectra computed with the bias parameter $b_\O=1$ for each $\O\in\O_3$.}
    \label{fig:triRSD} 
\end{figure} 

\vskip 4pt
Figure \ref{fig:triRSD} shows plots of the angular galaxy trispectrum from all cubic bias parameters in the equilateral configuration, using a Gaussian window function centered at redshift $z=1$ with $\sigma_z=0.1$. The trispectrum for $z=2$ looks very similar, with slightly lower amplitude and shifted scales, and hence we do not show it explicitly. Also not shown are the cross-terms between $b_\delta$, $b_{\delta^2}$, and $b_{\G_2}$. These are given in \eqref{tcross1}-\eqref{tcross3} and are straightforward to add. In generating the plots, we assumed constant bias parameters, and normalized the trispectrum by setting $b_\O=1$ for each bias and $b_\delta=1$ in all cases. The Limber approximation is used for both the external and internal multipoles, replacing $I_\ell^{(i)}$ and $J_L^{(s)}$ as in~\eqref{eq:limber}. We have chosen sufficiently large sampling points $N_r=100$ for the radial integrals, so that the results faithfully represent the true shapes, but not so large that the they have converged within 1\% accuracy over all multipoles. For instance, we find that the FFTLog method with $N_r=100$ agrees with a brute-force method of performing the numerical multi-dimensional integrals at the level of 1\% for low $\ell$'s, while the convergence is not reached for high $\ell$'s. As a consequence, we find a quantitative difference between the Limber and non-Limber results at high multipoles, which can differ up to a factor of two. However, we emphasize that this is only a numerical artifact, and the discrepancy indeed goes away upon increasing the sampling points up to e.g.~$N_r=400$. We will say more about the precision of the computation in the next section. Apart from this, the Limber approximation works well until it breaks down for small $\ell$, except for $b_{\G_2\delta}$ for which the approximation fails at almost all scales. The effect of RSD shows up at low $\ell$ as expected, but leads to a rather small amplitude difference.

\begin{figure}[t!]
    \centering
         \includegraphics[width=\textwidth]{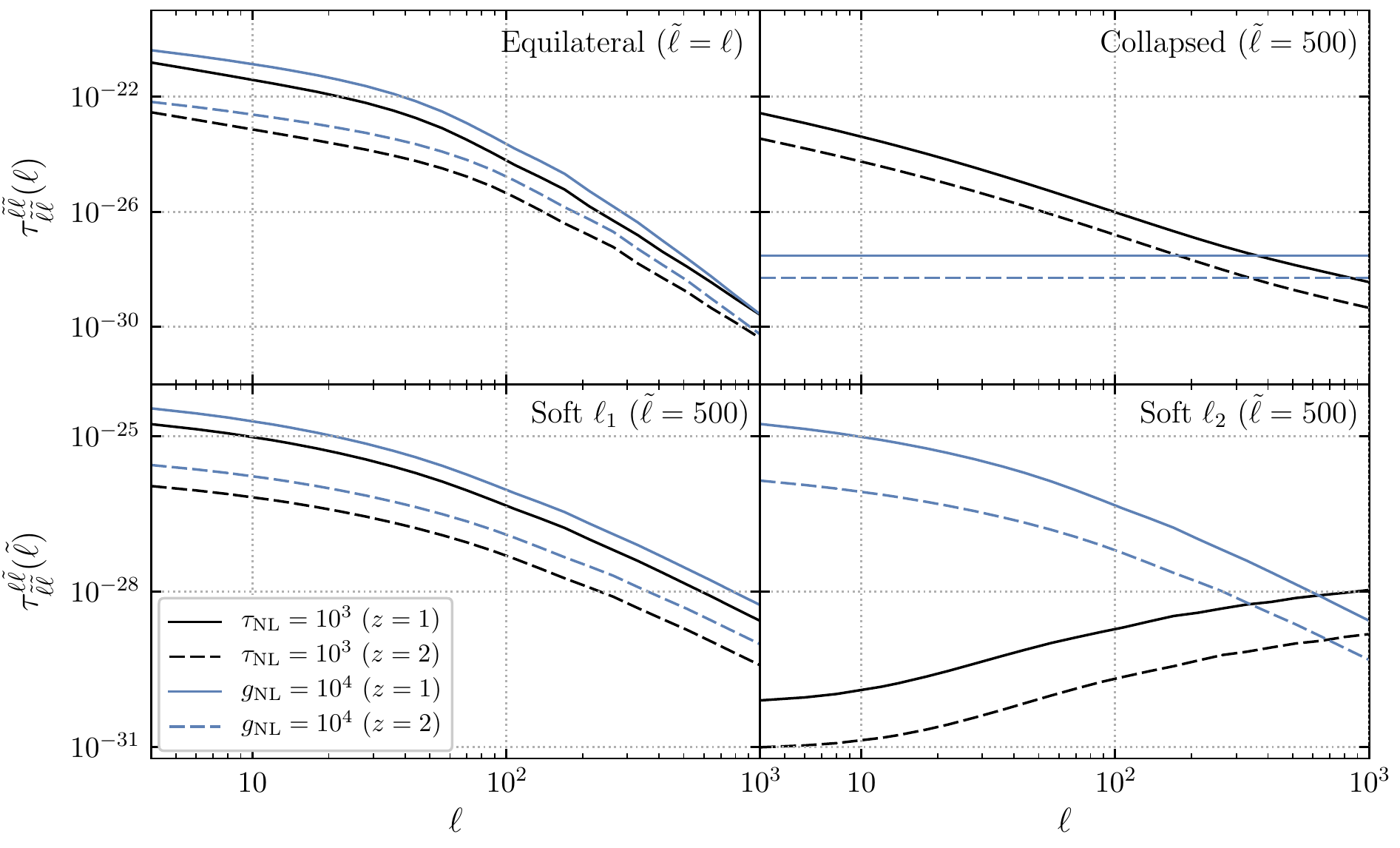}
    \caption{Angular galaxy trispectrum (with no RSD) from local primordial non-Gaussianity with $\tau_{\rm NL}=10^3$ (black) and $g_{\rm NL}=10^4$ (green). The solid and dashed lines are computed with Gaussian window functions centered at  $z=1,2$ with $\sigma_z=0.1,0.2$, respectively. In the upper panels, the left (right) plot shows the equilateral (collapsed) configurations. In the lower panels, the left (right) plot shows the soft $\ell_1$ ($\ell_2$) limit. We used $N_\eta=100$ terms in the FFTlog expansion of $\M(k)$ with $b=-1.1$.} 
    \label{fig:local} 
\end{figure}

\vskip 4pt
It is straightforward to compute the trispectrum with non-Gaussian initial conditions. For concreteness and simplicity, let us consider primordial non-Gaussianity of the local type. Figure~\ref{fig:local} shows plots of the tree-level angular galaxy trispectrum for two types of local non-Gaussianity---\eqref{tnlT} and \eqref{gnlT}---for different multipole configurations, with $\tnl=10^3$, $\gnl=10^4$, and $b_\delta=1$. To show the plot with multipole ranging up to $10^3$, we set the reference multipoles to $\tilde\ell=500$.  
In the two soft limits, the two trispectra behave differently: While they both grow as $\ell_1\to 0$ due to the presence of $P_\zeta(k_1)$ in the primordial trispectrum, only the $\gnl$ shape grows in the $\ell_2\to 0$ limit. This is somewhat misguiding, since it is an artifact of just looking at a single permutation; the full $\tnl$ trispectrum should grow in any $\ell_i\to 0$ limit. The difference between the shapes is instead most pronounced in the collapsed limit. We see that the $\tnl$ shape grows as $L\to 0$, while the $\gnl$ shape stays constant. This is easy to see from the primordial trispectrum, since $\gnl$ doesn't depend on the internal momentum. For equilateral configurations, both trispectra take similar shapes.

\subsection{Performance and Precision}\label{sec:perf}

An order of estimate for the computational cost of the angular trispectrum $\tau^{\ell_1\ell_2}_{\ell_3\ell_4}(L)$ in the scE-separable case for each multipole configuration is
\begin{align}
	{\cal N} = \O(N_\tau N_\eta N_r^2 N_\chi )\, ,
\end{align}
where $N_r$ and $N_\chi$ are the number of sampling points for numerically computing the $r$ and $\chi$ integrals, respectively, $N_\eta$ is the number of frequencies in the FFTLog decomposition, and $N_\tau$ is the number of separable terms in the trispectrum. Typically, $N_\chi\sim \O(50)$, $N_r\sim \O(100)$ and $N_\eta\sim\O(100)$ terms are required for convergence, whereas $N_\tau$ differs from term to term, and ranges between $1$ and $\O(100)$ depending on the number of terms in the bias operator considered. For the contact-separable case, the scaling reduces to ${\cal N} = O(N_\tau N_\eta N_r N_\chi )$.  We summarize the runtime and the parameters used for evaluating the angular galaxy trispectrum for a single configuration from different bias operators in Table~\ref{tab:performance}.

\vskip 5pt
\begin{table}[h!]
\begin{center}
\begin{tabular}{c c c c c c c c}
\tline\\[-10pt]
\hskip 5pt  Bias\phantom{\hskip 5pt} &\phantom{\hskip 2.5pt} Type \phantom{\hskip 2.5pt} & $N_\tau$ & $N_\eta$ & $N_r$ & $N_{r'}$  & $N_\chi$&\phantom{\hskip 5pt} $\tau^{\ell_1\ell_2}_{\ell_3\ell_4}(L)$ \phantom{\hskip 5pt} \ \\[3pt]
\tline\rowcolor[gray]{0.92}{} & & & & &  & & \\[-10pt]
	\rowcolor[gray]{0.92}{}$b_\delta$  & scE  & 76 & 200 & 100 &  100 &  50  & 2.5 min \\[5pt]
	\rowcolor[gray]{0.92}{}$b_{\delta^2}$ & scE  & 7 & 200 & 100 &  100 &  50  & 15 sec \\[5pt]
	\rowcolor[gray]{0.92}{}$b_{\delta^3}$ & C  & 1 & 200&100 & 0 & 50 & 1 sec\\[5pt]
	\rowcolor[gray]{0.92}{}$b_{\G_2}$ & scE & 72 & 200&100 & 100 & 50 & 2.5 min\\[5pt]
	\rowcolor[gray]{0.92}{}$b_{\G_2\delta}$ & C & 6 & 200&100 & 0 & 50 & 10 sec\\[5pt]
	\rowcolor[gray]{0.92}{}$b_{\Gamma_3}$ & scE & 36 & 200&100 & 100 & 50  & 1 min \\[5pt]
	\rowcolor[gray]{0.92}{}$b_{\G_3}$  & spE & 31 & 200&100 & 100 & 50  &  4 min\\[5pt]
\tline
\end{tabular}
\end{center}
\vspace{-0.5cm}
	\caption{Parameters and performance results for different bias operators. The last column denotes the approximate time for evaluating the trispectrum at a single multipole configuration using our \texttt{Mathematica} code on a laptop with Intel Core i9 CPU @ 2.3 GHz core.}
	\label{tab:performance}
\end{table}

\begin{figure}[t!]
    \centering
         \includegraphics[width=0.49\textwidth]{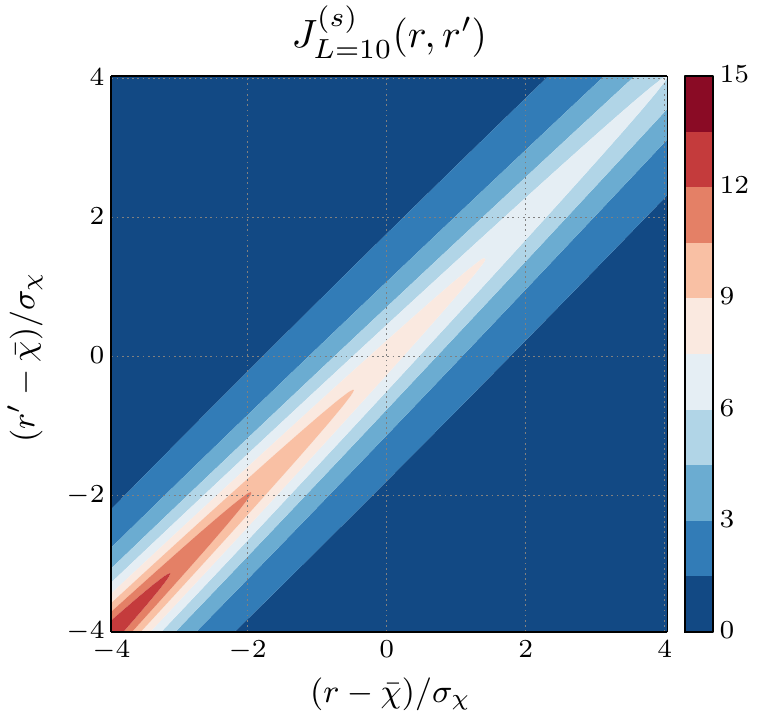}         \includegraphics[width=0.49\textwidth]{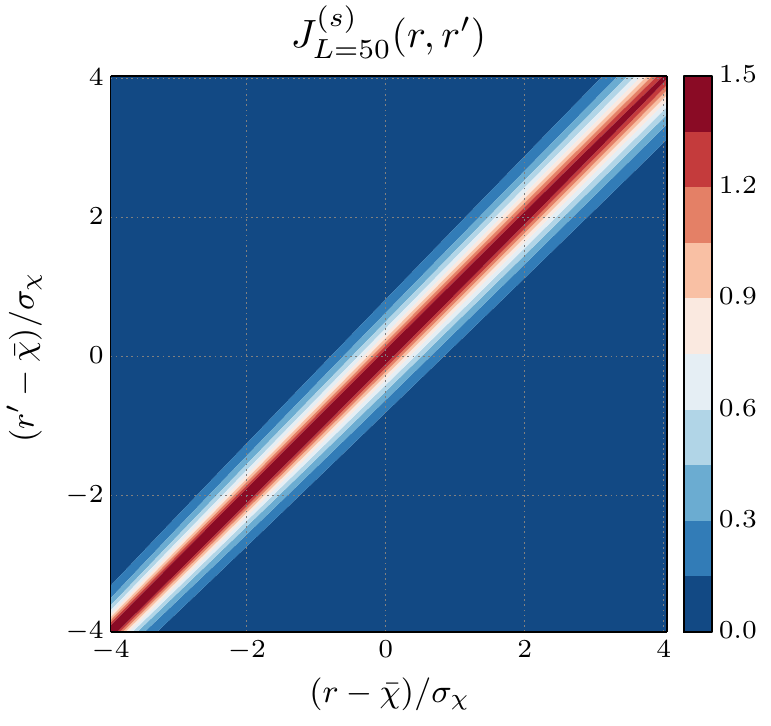}
    \caption{Contour plots of the coupling integral $J^{(s)}_L(r,r')$ with $p_s=-1$ for $L=10$ and $50$. The two axes represent the distance away from the point $(r,r')=(\bar\chi,\bar\chi)$ in units of $\sigma_\chi$, corresponding to $\bar z=1$ and $\sigma_z=0.1$.} 
    \label{fig:grid} 
\end{figure}

\vskip 4pt
Let us give some further remarks on our parameter choices. For exchange-separable trispectra, we chose $N_r= 100$ to generate the plots, which was enough to for the shapes to have sufficient converged over a wide range of (intermediate) multipoles. However, in order to reach convergence within 1\% accuracy over the entire multipole range, we require a much higher number of sampling points of about $N_r\sim 400$. This has to do with our current sampling scheme, in which we sample an $N_r\times N_r$ grid of equally-spaced points from the domain of a two-dimensional integral. In any numerical integration, the sampling scheme should be carefully chosen in order to faithfully represent the integral. In our one-dimensional problem, the $\chi$-integrand of $I_\ell^{(i)}(r)$ is dictated by the Gaussian window function, which is peaked at $\chi=\bar\chi$ with a width $\sigma_\chi$. We find that about $N_\chi\sim 50$ is enough to sufficiently sample this integral. The resulting function $I_\ell^{(i)}(r)$ then inherits this shape, which is also peaked around $r=\bar\chi$ with the same width, as shown in Fig.~\ref{fig:RSDint}. 
Naively, this suggests that the $(r,r')$-integrand should follow a bivariate Gaussian shape centered at $r=r'=\bar\chi$ with radius $\sigma_\chi$ (for equal redshifts). This would be true if the two integrals are factorized, but is obviously false in our problem due to the presence of the coupling integral $J_L^{(s)}(r,r')$. The integral has no knowledge about $\bar\chi$, and is instead peak at $r=r'$. In Fig.~\ref{fig:grid}, we show the behavior of the coupling integral in the two-dimensional plane around the center $(r,r')=(\bar\chi,\bar\chi)$ in units of $\sigma_\chi$, corresponding to our canonical choice $\bar z=1$ and $\sigma_z=0.1$. We see that the two-dimensional integral is dominated along the line $r=r'$ and quickly diminishes away from it. This behavior becomes more extreme for higher $L$, which implies that sampling for the double radial integral can be done almost one-dimensionally. Indeed, the Limber approximation is the limit in which the domain of integration precisely reduces to the line $r=r'$. We have not optimized the integration scheme, and simply used a square grid of points to sample the integral, implemented as an $N_r\times N_r$ matrix multiplication. This can be rather cost-ineffective for large $N_r$, and a better sampling scheme can be implemented to effectively reduce the computational cost of the radial integration for high $\ell$.\footnote{In the statistics context, there is a well-known method to efficiently sample correlated variables with multivariate normal distributions using the Cholesky decomposition of the covariance matrix. Our problem is similar, with the coupling integral playing the role of the covariance matrix with a strong positive correlation.}

	\vskip 4pt
For a given redshift, an angular trispectrum is described by five independent degrees of freedom. Using $N$ $\ell$-bins, there would be a total of $\O(N^5)$ trispectrum configurations to be evaluated, which gets reduced by a factor of $4!$ due to permutation symmetry. Since it takes about $\O(1)$ minutes to evaluate the trispectrum for a single configuration (see  Table~\ref{tab:performance}), using $N=10$ we would require about $\O(10^2)$ CPU hours to compute the trispectrum for all configurations, for a fixed cosmology and redshift. As discussed above, we have not optimized our integration method in our code, and a better performance can be achieved with parallelization and an improved sampling scheme.

\section{Non-Gaussian Covariance of Angular Power Spectrum}\label{sec:cov}

In order to obtain accurate constraints on cosmological parameters, it is important to have a precise theoretical prediction for the covariance. However, to the best of our knowledge, the full computation of the covariance matrix for the power spectrum (including the connected part) in angular space beyond the Limber approximation has not yet been performed, due to the difficulty associated with computing the trispectrum. 
In this section, equipped with the formalism for computing the angular trispectrum, we compute the non-Gaussian covariance of the angular power spectrum from the connected part of the trispectrum.\footnote{In the literature, other types of the power spectrum covariance in multipole space have also been studied. For example, one may consider a partial wave expansion of the anisotropic power spectrum with respect to the angle between the momentum vector and a line-of-sight direction, relevant for the galaxy power spectrum with RSD~\cite{Sugiyama:2019ike, Wadekar:2019rdu}. The resulting covariance depends both on the multipole and the wavenumber $k$, with nonzero monopole, quadrupole and octupole. This is clearly different from the object we are computing, which is the covariance of the angular power spectrum in $\ell m$-space.} For simplicity, we set $b_\delta=1$ and all other bias parameters to zero, which is equivalent to computing the matter power spectrum covariance.

\subsection{Power Spectrum Estimator and Covariance}
At tree level, the theoretical angular matter power spectrum can be computed as
\begin{align}
	C_\ell = \frac{1}{2\pi^2}\int_0^\infty \d r\, \W_\delta(r) I_\ell(r)\, ,
\end{align}
where $I_\ell(r)$ is defined as in \eqref{Ifunc} with $f_i(k,z)=D_g(z)P(k)$. Since the power spectrum only consists of a single radial integral, the sampling points do not need to be as dense as in the trispectrum calculation, and $N_r=50$ is sufficient. The optimal estimator $\widehat C_\ell$ for the angular power spectrum is given by summing over all measured multipoles,
\begin{align}
	\widehat C_\ell = \frac{1}{2\ell+1} \sum_{m=-\ell}^\ell \big|\delta_{\ell m}^{(\rm obs)}\big|^2\, ,
\end{align}
which is unbiased as $\langle\widehat C_\ell\rangle = C_\ell$.\footnote{In practice, the observable is not measured over the entire sky. This means that different multipoles in the spherical harmonic expansion will be correlated, which results in a biased estimator. The effect of partial sky coverage can be accounted for by adding a mode-coupling kernel~\cite{Hivon:2001jp}.} 
The covariance matrix of the power spectrum estimator is~\cite{Hu:2001fa}
\begin{align}
	{\sf C}_{\ell\ell'} &= \frac{(-1)^{\ell+\ell'}}{\sqrt{(2\ell+1)(2\ell'+1)}}\,T^{\ell\ell}_{\ell'\ell'}(0) - C_{\ell}C_{\ell'} \label{Covell}\\
	&=\frac{2\delta_{\ell\ell'}}{2\ell+1}\,C_{\ell}^2+\frac{(-1)^{\ell+\ell'}}{\sqrt{(2\ell+1)(2\ell'+1)}}\left[P^{\ell\ell}_{\ell'\ell'}(0)+\frac{2}{\sqrt{(2\ell+1)(2\ell'+1)}}\sum_L (-1)^L P^{\ell\ell}_{\ell'\ell'}(L)\right]\,,\nonumber
\end{align}
where in the second line we have subtracted the purely disconnected piece from the trispectrum. The trispectrum in a single channel $P^{\ell\ell}_{\ell'\ell'}(L)$ can be obtained by summing over permutations of reduced trispectra as in \eqref{Pdecomp}. The symmetries of $P^{\ell\ell}_{\ell'\ell'}(L)$ implies that it vanishes for odd $L$, so the $L$-sum is only over even multipoles. This enters in the Fisher matrix as usual,
\begin{align}
	F_{\alpha\beta} = \sum_{\ell\ell'}\frac{\partial C_\ell}{\partial \lambda_\alpha}\hs{\sf C}_{\ell\ell'}^{-1}\hs\frac{\partial C_{\ell'}}{\partial \lambda_\beta}\, ,
\end{align}
for a set of parameters $\{\lambda_\alpha\}$. The diagonal components of the Fisher matrix give the marginalized 1-$\sigma$ uncertainties in the parameters $\lambda_\alpha$. The off-diagonal components from the non-Gaussian part induce correlations between the uncertainties, which in general degrade parameter constraints. 

\vskip 4pt
It is instructive to compare \eqref{Covell} to the analogous calculation in Fourier space. The power spectrum estimator in Fourier space is
\begin{align}
	\widehat P(k) = V_f \int_{V_s(k)} \frac{\d^3 q}{V_s(k)} \delta(\q)\delta(-\q)\, ,
\end{align}
where $V_f=(2\pi)^3/V$ is the volume of the fundamental shell and the integration is performed over the is the differential volume of the shell of radius $k$, $V_s(k)=4\pi k^2\delta k$. 
The covariance is given by~\cite{Scoccimarro:1999kp} (see also~\cite{Mohammed:2016sre, Barreira:2017kxd})
\begin{align}
	{\sf C}(k,k') &= \frac{P(k)^2}{(V_s(k)/V_f)}\delta_{kk'}+ \frac{1}{V}\int_{V_s(k)}\frac{\d^3 q}{V_s(k)}\int_{V_s(k)}\frac{\d^3 q'}{V_s(k')}\,T(\q,-\q,\q',-\q')\, .
\end{align}
We see that the covariance receives contribution from the trispectrum only in the {\it collapsed} configuration, which corresponds to the limit in which the internal momentum is collapsed to zero length, i.e.~$s\to 0$ in the $s$-channel. 
Note that the covariance in angular space \eqref{Covell}, too, is evaluated in the collapsed multipole configuration $L=0$. When projected on the sphere, we are integrating over all momenta, so the covariance in some sense receives contributions from all wavelengths. However, the covariance in angular space is still mostly captured by terms that dominate in the $s\to 0$ limit.

\subsection{Angular Matter Power Spectrum Covariance at Tree Level}

We now turn to the computation of the non-Gaussian covariance of the angular matter power spectrum. There are essentially three most relevant pieces that contribute to the non-Gaussian covariance: the connected four-point function at tree level and one loop, and the super-sample covariance~\cite{Takada:2013bfn, Li:2014sga}. In this section, we consider the contribution from the tree-level trispectrum. 

\begin{figure}[t!]
    \centering
         \includegraphics[height=6.5cm]{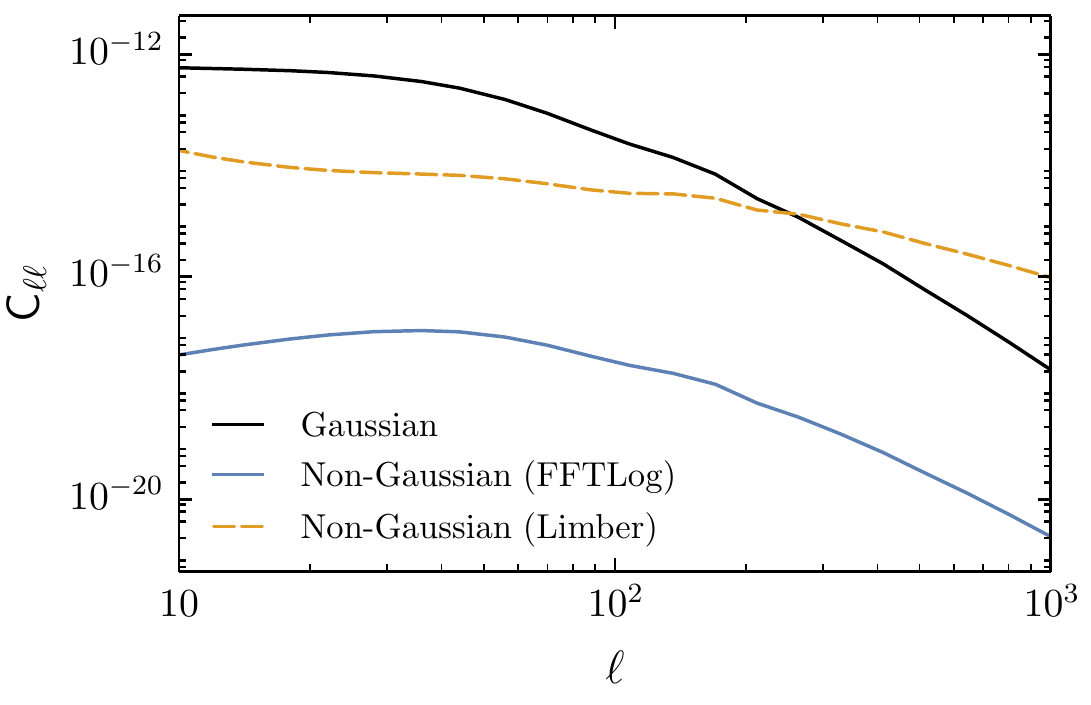} \\
    \caption{Comparison of Gaussian and non-Gaussian covariance of the angular matter power spectrum at $z=1$ with $\sigma_z=0.1$. Two results are shown for the non-Gaussian covariance, one computed with the FFTLog method and other using the Limber approximation. The latter is negative, so its absolute value is shown. As explained in the main text, the Limber-based result is unphysical.}
    \label{fig:cov}
\end{figure}

\vskip 4pt
Figure~\ref{fig:cov} shows the diagonal elements of the covariance from the Gaussian and non-Gaussian parts. We show two results for the latter part, one using the FFTLog method and the other with the Limber approximation. Since we are evaluating the trispectrum at $L=0$, this time we use the Limber approximation only for the external multipoles, and the coupling integral $J_L^{(s)}$ was computed with the FFTLog method in both cases. One can immediately notice that there is quite a large discrepancy between the FFTLog- and Limber-based calculations, even for large multipoles for which we normally think that the Limber approximation should be valid. We argue that this Limber-based calculation cannot be trusted. This has to do with the subtle issue about the numerical accuracy of the Limber approximation, as we explain further below. The FFTLog calculation shows that the non-Gaussian part from the connected four-point function gives a small contribution to the full covariance in angular space.

\subsubsection*{FFTLog vs. Limber}
First of all, we would like to know which terms in the trispectrum cause the large discrepancy between the FFTLog method and Limber approximation. 
To understand the root of the problem, we decompose the terms proportional to $s^{-4}$ in $T_{2211}$, which give a large contribution to the covariance. From \eqref{T2211g}, we have (see also Appendix~\ref{app:div})
\begin{align}
	T_{2211} \supset\ D_1D_2^2D_3D_4^2P(k_1)P(k_3)P(s)\left(3k_2^2 -5k_1^2+\frac{2k_2^4}{k_1^2}\right)\left(3k_4^2 -5k_3^2+\frac{2k_4^4}{k_3^2}\right)\frac{1}{s^4}\, .
\end{align}
This leads to the radial integral of the form
\begin{align}
	\int_0^\infty \d r\, r^2\big(\underbrace{3I_\ell^{(1,0)}I_\ell^{(2,1)}}_{e_1} +\underbrace{(-\, 5I_\ell^{(1,1)}I_\ell^{(2,0)})}_{e_2}+\underbrace{2I_\ell^{(1,-1)}I_\ell^{(2,2)}}_{e_3}\big)J_0^{(s,-4)},\label{rintegrand0}
\end{align}
where we have suppressed the arguments and the second radial integral consisting of terms depending on $k_3$, $k_4$. The issue is that the values of all three terms in \eqref{rintegrand0} are quite similar, with cancellations occurring at $10^{-5}$ level. We thus need to evaluate $I_\ell^{(i)}$ at very high precision in order to account for the correct cancellation between these terms. Note that this cancellation occurs already at the level of the $r$ integrand, so we can see that this is problematic even before computing the full trispectrum, and that this is unrelated to the sampling scheme we choose for the radial integrals. In Table~\ref{tab:limber}, we tabulate the values of the terms proportional to $s^{-4}$, computed with and without the Limber approximation. We see that although individual terms computed using the Limber approximation agree with the FFTLog-based result at 1\% level, this accuracy is not enough to ensure the full cancellation between the terms, resulting in an orders of magnitude difference in the final trispectrum. We find a similar level of cancellation occurs for the terms proportional to $s^{-2}$, whereas the Limber approximation works well for those proportional to $s^0$, $s^2$, and $s^4$.

\vskip 5pt
\begin{table}[h!]
\begin{center}
\begin{tabular}{ccccc}
\tline\\[-10pt]
\hskip 10pt $\tau^{\ell\ell}_{\ell\ell}(0)$ $(\times 10^{10})$\phantom{\hskip 5pt} & $e_1$ \phantom{\hskip 5pt} & \phantom{\hskip 5pt}$e_2$\phantom{\hskip 5pt}  &\phantom{\hskip 5pt} $e_3$ \phantom{\hskip 5pt}&\phantom{\hskip 5pt} $e_1+e_2+e_3$\phantom{\hskip 10pt} \ \\[3pt]
\tline\rowcolor[gray]{0.92}{} & &  & & \\[-10pt]
	\rowcolor[gray]{0.92}{}FFTLog &  $-10.3936$ &  $17.3228$ &  $-6.92884$ & $3.80825\times 10^{-4}$ \\[5pt]
	\rowcolor[gray]{0.92}{} & &  & & \\[-10pt]
	\rowcolor[gray]{0.92}{}Limber  & $-10.4896$& $17.4829$ & $-7.01109$ & $-1.77898\times 10^{-2}$ \\[5pt]
\tline
\end{tabular}
\end{center}
\vspace{-0.5cm}
	\caption{Comparison of angular matter trispectrum computed using the FFTLog method and the Limber approximation from terms proportional to $s^{-4}$ in $T_{2211}$, evaluated at $\ell=100$, $L=0$.}
	\label{tab:limber}
\end{table}

\begin{figure}[t!]
    \centering
                 \includegraphics[width=\textwidth]{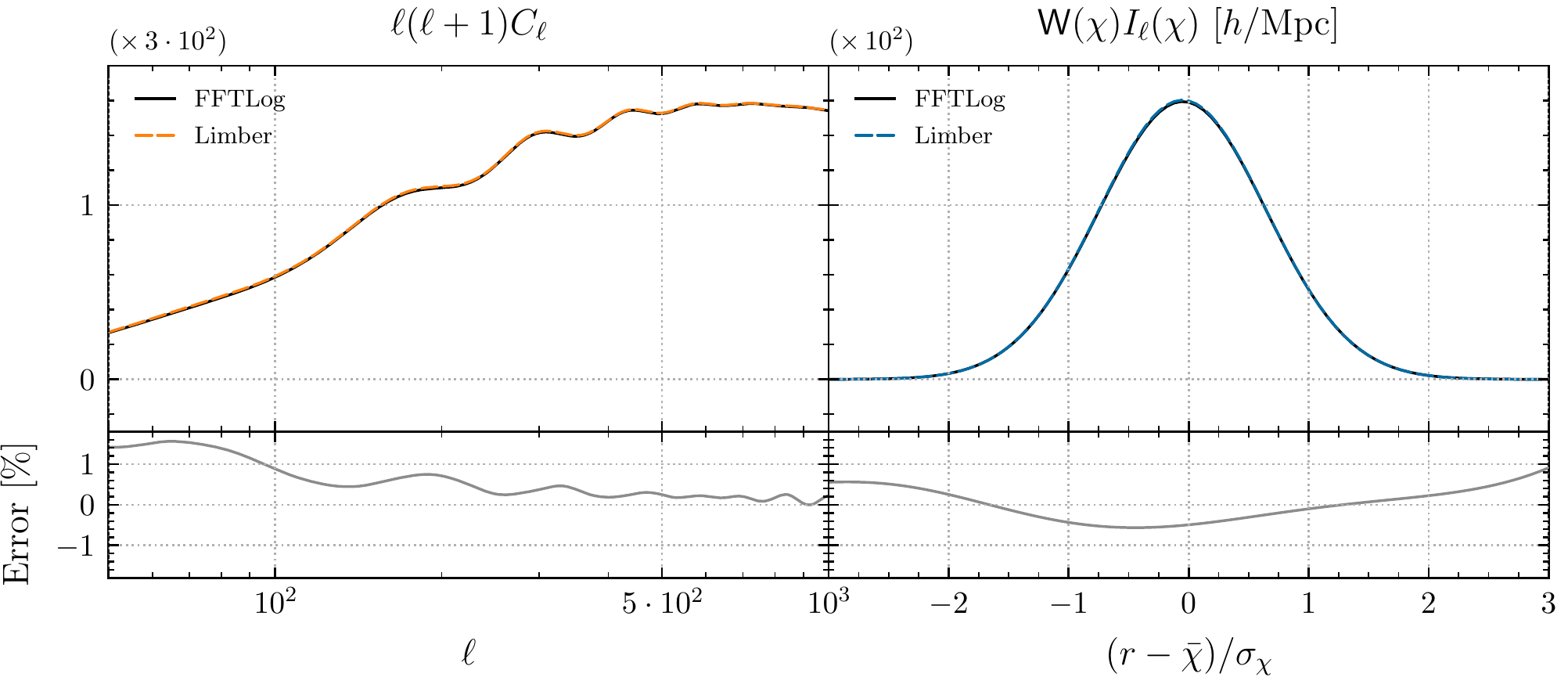}          
    \caption{Comparison between the FFTLog method and the Limber approximation for the angular matter power spectrum ({\it left}) and its integrand for $\ell=300$ ({\it right}). The bottom panel for each plot shows the relative error in percentage.  
   We used the window function located at $z=1$ with $\sigma_z=0.1$, and chose higher number of sampling points $N_r=N_\chi=200$ than usual.}
    \label{fig:CellLimber}
\end{figure}

A desired level of precision may be achieved with the FFTLog method by choosing a sufficiently large number of sampling points. In contrast, the Limber approximation has an intrinsic level of error, simply due to the fact that it replaces the Bessel function with a Dirac delta function, whose accuracy also depends on the width of the window function used. We find that the level of accuracy of the Limber approximation is roughly at $0.1\%$ level for large multipoles. This can be seen from e.g.~Fig.~\ref{fig:CellLimber}, where we show the comparison between the FFTLog method and the Limber approximation, both for the angular matter power spectrum and its radial integrand. In producing the plots, we used high enough precision to make sure that the results converged for each Limber and non-Limber calculation.

\vskip 4pt
It is rather striking that the Limber approximation can dramatically fail in a range of scales we normally think it can be safely trusted. We find that increasing the precision of numerical integrations does not change this conclusion. As far as we are aware, a similar observation have not been made for lower-point functions, likely because the kinematic configurations of two- and three-point functions are simpler than the four-point case. 
The validity of the Limber approximation should thus be carefully checked whenever it is used, especially when dealing with the non-Gaussian covariance.

\section{Conclusions}\label{sec:con}

In the era of high-precision cosmology and large datasets, it is important to build efficient algorithms for calculating and estimating cosmological observables. 
In this paper, we presented an efficient semi-analytic method to compute cosmological angular trispectra. This generalizes the method of~\cite{Assassi:2017lea} to four-point angular statistics, and we used the method to compute the galaxy angular trispectrum and the non-Gaussian covariance of the angular matter power spectrum. We also defined a suitable separable ansatz for cosmological four-point functions, and classified their separability types based on the physical criteria that correlators ought to satisfy. 

\vskip 4pt
There are numerous other applications in cosmology one could explore using the FFTLog algorithm. First of all, it would be interesting to generalize the current formalism to other types of angular observables. Some would be simpler than others: The trivial list involves angular trispectra of the cosmic microwave background, weak lensing, etc., which simply require modifications of the line-of-sight integral kernels. Similarly, tensor observables in angular space, carrying the same Bessel integral structure as spin-0 fields but dressed with more complicated geometric dependence, involve a straightforward generalization. 

\vskip 4pt
More nontrivial applications involve applying the formalism to observables that go beyond the linear regime. In perturbation theory, these are systematically captured by loop integrals in Fourier space. A parallel investigation of FFTLog-based methods in Fourier space~\cite{Schmittfull:2016jsw, Schmittfull:2016yqx, Simonovic:2017mhp, Slepian:2018vds} has revealed that power-law cosmologies have many analytic solutions in this case too, bearing similarities with standard loop integrals in quantum field theory. It is natural to unify the two methods to compute e.g. the one-loop angular power spectrum. There also exist more phenomenological approaches to nonlinear scales in the large-scale structure such as the halo model~\cite{Cooray:2002dia}. In this setup, one deals with extra layers of integrals over halo profiles to compute correlators, but otherwise whose integrands consist of separable products of functions. It should thus be possible in principle to apply the FFTLog algorithm to the halo model as well, both in angular and in Fourier space.

\vskip 4pt
Another interesting application involves building efficient separable templates for inflationary correlation functions. Many shapes can be made separable using the trick~\eqref{kt}, and other traditional approaches involving estimating shapes in terms of an orthogonal basis of polynomials~\cite{Regan:2010cn, Fergusson:2010ia, Schmittfull:2012hq}. However, sometimes it is possible to exploit the soft limit behavior of the correlators to directly construct separable templates by pure ansatz, see e.g.~\cite{Creminelli:2006rz, MoradinezhadDizgah:2019xun}. 
In particular, it would be nice if some simple templates for the equilateral and exchange-type four-point functions can be constructed in this way. 
We have not fully explored the shape dependence of the angular galaxy trispectrum in this work. Having easier way to compute primordial trispectra in angular space would also make it possible to study the correlations between different shapes. For four-point functions, there is a reduction in the number of degrees of freedom of shapes when going from three to two dimensions, so it is not totally obvious whether the shape correlations will remain the same after projection. It would be interesting to systematically study correlations between non-Gaussian shapes from inflation and from bias parameters, and run a detailed forecast to figure out which shapes are easier to detect than others in the large-scale structure.

\vskip 4pt
One of the main utilities of our formalism is a fast and reliable way of computing the non-Gaussian covariance of the power spectrum. 
Usually, the Limber approximation is believed to work well in the small-scale regime. However, our investigations show a direct counter example. This happens in situations where there are numerical cancellations that are finer than the intrinsic level of precision that the Limber approximation offers, which we find to be about $O(0.1)\%$ at high $\ell$. 
It would be nice to explore the viability of the Limber approximation more generally, as well as that of the closely related flat-sky approximation, 
and how this affects cosmological parameter estimation in future surveys (see~\cite{Kitching:2016zkn, Kilbinger:2017lvu, Lemos:2017arq, Fang:2019xat} for recent works). Our formalism paves a way towards quantitatively tackling the question about how important the non-Gaussian contribution is to the covariance in angular space. Some initial steps in this direction were taken recently in~\cite{Lacasa:2017ufk, Lacasa:2018hqp, Lacasa:2019orz, Lacasa:2019flz}, mainly focusing on a subset of terms that are amenable to the Limber approximation in the halo model. We leave further progress in this interesting direction to future work.

\paragraph{Acknowledgement} We thank Azadeh Moradinezhad Dizgah for initial collaboration, and Daniel Eisenstein, Julian Mu\~noz, and Matias Zaldarriaga for useful discussions. CD and HL are partially supported by Department of Energy (DOE) grant DE-SC0020223. 

\newpage
\appendix

\section{Spurious Divergence}\label{app:div}

In computing angular correlations, we convert the momentum-conserving delta functions into integrals over plane waves. These integrals are numerically evaluated after performing the momentum integrals. Because of this, the momentum integrals may give rise to spurious divergences from unphysical momentum configurations in the intermediate steps. For the exchange-separable trispectrum, this happens when the coupling integral
\begin{align}
	J_L^{(s)}(r,r') = 4\pi\sum_m c_m\int_0^\infty\d s\hs s^{\nu_m+2p_s-1}j_L(sr)j_L(sr')\, ,
\end{align}
is divergent, where $\nu_m=3-b+i\eta_m$. 
The (non-)divergent nature of this integral is determined by the integer $p_s$: for $1<b<2$, it is UV divergent for $p_s>0$ and IR divergent for $p_s< -L$.
In this appendix, we describe a procedure to remove these spurious divergences that were encountered in the main text.

\subsection{IR Divergence}

Let us first consider the IR divergence, which occurs when $p_s<-L$. For the matter trispectrum, we saw that the minimum of $p_s$ was given by $p_s=-2$, making it divergent for $L=0,1$. We will consider the $L=0$ case here, since this is what is relevant for computing the covariance from the connected trispectrum. 

\vskip 4pt 
To see how this works in practice, we consider the naively divergent term in the matter trispectrum,
\begin{align}
	T_{2211} \supset\ D_1D_2^2D_3D_4^2P(k_1)P(k_3)P(s)\left(3k_2^2 -5k_1^2+\frac{2k_2^4}{k_1^2}\right)\left(3k_4^2 -5k_3^2+\frac{2k_4^4}{k_3^2}\right)\frac{1}{s^4}\, ,
\end{align}
where we dropped a constant prefactor. To remove this apparent singularity in our integral formulas, we will soften the UV behavior of the terms inside the brackets by integrating by parts. As we saw in \eqref{eq:Iellr3}, we remove such UV divergence of $k_i$ by shifting the frequency with the action of $\D_\ell(\chi)$ on $\I_\ell$ and then integrating it by parts to act it acts on the window function. This time, we instead shift the frequency with the operator $\D_\ell(r)$ and then integrate by part to hit the coupling integral $J^{(s)}_L(r,r')$. 

\vskip 4pt
Let us look at the 1,2-leg first. We label each integral with $I_{\ell_i}^{(i,p_i)}$ and $J_L^{(s,p_s)}$. For $L=0$, we must have $\ell_1=\ell_2\equiv\ell$. Then the $r$-integral consists of terms
\begin{align}
	\int_0^\infty \d r\, r^2\big(3I_\ell^{(1,0)}I_\ell^{(2,1)}-5I_\ell^{(1,1)}I_\ell^{(2,0)}+2I_\ell^{(1,-1)}I_\ell^{(2,2)}\big)J_0^{(s,-4)}\, ,\label{rintegrand}
\end{align}
where we have suppressed the arguments. Let us first shift $I_{\ell}^{2,1}= \D_{\ell}I_{\ell}^{1,0}$ in the first term. We then get
\begin{align}
	3r^2 I_{\ell}^{(1,0)}I_{\ell}^{(2,1)}J_0^{(s,-4)}\, \to\, 3r^2 I_{\ell}^{(2,0)}\big(I_{\ell}^{(1,1)}J_0^{(s,-4)}+I_{\ell}^{(1,0)}J_0^{(s,-2)}+2\partial_r I_{\ell}^{(1,0)}\partial_r J_0^{(s,-4)}\big)\, .
\end{align}
We can disregard the boundary terms in this process, since $I_{\ell}^{(2,p_2)}$ essentially behaves as a window function, e.g.~$I_{\ell}^{(2,p_2)}=\frac{2\pi^2}{r^2}\D_\ell^{p_2}\W^{(2)}$ for $p_2\ge 0$, and thus have zero support on the boundary. 
Now, we integrate by parts twice the last term in \eqref{rintegrand} to shift $I_\ell^{(2,2)}\to I_\ell^{(2,0)}$. Doing so, the leading divergent piece precisely cancels with other terms, and the $r$ integral becomes
\begin{equation}
	\hskip -12pt\int_0^\infty\! \d r\hs r^2\Big[\big(5I_\ell^{(1,0)}I_\ell^{(2,0)}{+}2I_\ell^{(1,-1)}I_\ell^{(2,1)}\big)J_0^{(-2)}{-2}\big(2I_\ell^{(2,0)}\partial_r I_\ell^{(1,-1)}{+}5I_\ell^{(2,0)}\partial_r I_\ell^{(1,0)}\big)\partial_r J_0^{(-4)}\Big] .
\end{equation}
By using the identity $\partial_r j_\ell(kr) = \frac{\ell}{r}j_\ell(kr)-k j_{\ell+1}(kr)$, the radial derivative of $I_\ell^{(i,p_i)}(r)$ can be expressed as
\begin{align}
	\partial_r I_\ell^{(i,p_i)}(r) &= \sum_n c_n r^{-1-\nu_n} \int_0^\infty \d\chi\, \Big[(1-\nu)\W_\delta(\chi) + \chi \W_\delta'(\chi)\Big] \I_\ell(\nu_n+2p_i,\tfrac{\chi}{r})\, ,
\end{align}
i.e.~the action of $\partial_r$ leads to a modification of the window function. Similarly, the derivative of $\partial_r J_0^{(-4)}$ can be written as
\begin{align}
	\partial_r J_0^{(s,-4)}(r,r') 
	&= -\sum_n c_n\, \int_0^\infty \d k\, k^{\nu_n+2p_s}j_1(kr)j_0(kr')\, .
\end{align}
Since $j_1(kr)\sim k$ as $k\to 0$, this has the same degree of divergence as $J_0^{(s,-2)}$. We can go through the same exercise for the 3,4-legs, with $\ell_3=\ell_4=\ell'$. This will lead to terms such as $J_0^{(s,0)}$, $\partial_r J_0^{(s,-2)}$, $\partial_{r'} J_0^{(s,-2)}$, and $\partial_r\partial_{r'}J_0^{(s,-4)}$, which are manifestly free of IR divergences. 

\subsection{UV Divergence}

Dealing with UV divergences is simpler than the IR divergent case. As we just saw, in the latter case we have to make sure the divergent $J_L^{(s)}$ terms cancel off each other after integration by parts. In the former case, we simply need to integrate by parts a sufficient number of times to remove the divergence of $J_L^{(s)}$. 
The way it works is that integrating by parts shifts UV divergences of $J_L^{(s)}$ to $I_{\ell_i}^{(i)}$, which we can handle using the trick introduced in \eqref{eq:Iellr3}. For instance, we can lower the frequency $p_s$ by one unit by writing $J_L^{(s,p_s)}=\D_L\, J_L^{(s,p_s-1)}$ and then integrating by parts as
\begin{align}
	\int_0^\infty &\d r\, r^2 I_{\ell_1}^{(1,p_1)}(r)I_{\ell_2}^{(2,p_2)}(r)J_L^{(s,p_s)}(r,r') = \int_0^\infty\d r\, J_L^{(s,p_s-1)}(r,r')\tilde\D_L(r) \Big[r^{2}I_{\ell_1}^{(1,p_1)}(r)I_{\ell_2}^{(2,p_2)}(r)\Big] \nn
	&= \int_0^\infty\d r\, J_L^{(s,p_s-1)}(r,r')\Big[\big(L(L+1)-\ell_1(\ell_1+1)-\ell_2(\ell_1+1)\big)I_{\ell_1}^{(1,p_1)}(r)I_{\ell_2}^{(2,p_2)}(r)\nn[5pt]
	& \quad + r^2 \big(I_{\ell_1}^{(1,p_1+1)}(r) I_{\ell_2}^{(2,p_2)}(r)+ I_{\ell_1}^{(1,p_1)}(r) I_{\ell_2}^{(2,p_2+1)}(r)-2\partial_r I_{\ell_1}^{(1,p_1)}(r)\partial_r I_{\ell_2}^{(2,p_2)}(r)\big)\Big]\, ,
\end{align}
where we have only shown the $r$ integral and written $\tilde \D_L$ in terms of $\D_{\ell_i}$ that in turn shift the frequencies of $I_{\ell_i}^{(i,p_i)}$. Since the matter trispectrum contains terms up to $p_s=2$, we would need to integrate by parts twice. This can get quickly complicated, giving rise to many terms and thus slowing down the computation. Since the radial integrands are rather smooth, in practice we can also take these derivatives numerically. To avoid numerical instability of taking multiple numerical differentiation, one should apply derivatives on the $r$ and $r'$ integrands as symmetrically as possible.

\section{Cubic Bias}\label{app:cubic}
In this appendix, we provide explicit expressions of the galaxy trispectrum in momentum space in \S\ref{sec:bias} in terms of the variables $\{k_1,k_2,k_3,k_4,s,t\}$ arising from the set $\O_3$ of all independent cubic bias operators listed in \eqref{O3}.  The set of operators we used is of course not a unique choice. Another common choice of bias operators employed in the literature is (see e.g.~\cite{Desjacques:2016bnm})
\begin{equation}
	\big\{ \delta,\, \delta^2,\, K^2,\, \delta^3,\, K^2\delta,\, K^3,\, O_{\rm td} \big\} \, .
\end{equation}
The two sets of operators are related by
\begin{align}
	\begin{bmatrix}
	\delta \\ \delta^2 \\ \G_2 \\ \delta^3 \\ \G_2\delta \\ \G_3 \\ \Gamma_3	
	\end{bmatrix} =
\begin{bmatrix}
		1 & 0 & 0 & 0 & 0 & 0 & 0 \\ 
		0 & 1 & 0 & 0 & 0 & 0 & 0 \\ 
		0 & -\frac{2}{3} & 1 & 0 & 0 & 0 & 0 \\
		0 & 0 & 0 & 1 & 0 & 0 & 0 \\
		0 & 0 & 0 & -\frac{2}{3} & 1 & 0 & 0 \\
		0 & 0 & 0 & -\frac{1}{9} & \frac{1}{2} & -1 & 0 \\
		0 & 0 & 0 & -\frac{16}{63} & \frac{8}{21} & 0 & 1
	\end{bmatrix}
	\begin{bmatrix}
	\delta \\ \delta^2 \\ K^2 \\ \delta^3 \\ K^2\delta \\ K^3 \\ O_{\rm td}
	\end{bmatrix} \, .
\end{align}
More relations between third-order bias parameters can be found in~\cite{Angulo:2015eqa, Lazeyras:2017hxw, Abidi:2018eyd}.

\vskip 4pt
Let us summarize the momentum dependence of the $T_{3311}$ part of the galaxy trispectrum in arising from different bias parameters $b_\O$, denoted by $F_\O$. (For the $T_{2211}$ part, see \eqref{F2symT2211}.) In the $s$-channel, they are given by
\begin{align}
	F_{\G_2\delta}(\k_1,\k_2) &\equiv \sigma_{\k_1,\k_2}^2\, ,\\
	F_{\delta^2}(\k_1,\k_2) &\equiv F_2^{\rm sym}(\k_1,\k_2)\, ,\\
	F_{\delta^3}(\k_1,\k_2,\k_3) &\equiv F_3^{\rm sym}(\k_1,\k_2,\k_3)\, ,\\
	F_{\G_3}(\k_1,\k_2,\k_3) &\equiv 2(\hat\k_1\cdot\hat\k_2)(\hat\k_1\cdot\hat\k_3)(\hat\k_2\cdot\hat\k_3)+\sigma_{\k_1,\k_2}^2\, ,\\
	F_{\Gamma_3}(\k_1,\k_2,\k_3) &\equiv \sigma_{\k_1,\k_{23}}^2\big(F_2^{\rm sym}(\k_1,\k_2)-G_2^{\rm sym}(\k_1,\k_2)\big)\, .
\end{align}
When expressed in terms of the scalar variables, these become rather lengthy expressions, so we introduce a compact notation
\begin{align}
	K^{(m,n)}  &\equiv \sum_{i\ne j}^3 k_i^m k_j^n\, ,\quad k_{ij,\pm }^{(m,n)} \equiv (k_i^m\pm k_j^m)^n\, ,
\end{align}
to express the external wavenumber dependence. We have
\begin{align}
	\sigma_{\k_1,\k_2}^2 &= \frac{k_{12,-}^{(1,2)}k_{12,+}^{(1,2)}-2k_{12,+}^{(2,1)}+s^4}{4k_1^2k_2^2}\, , \\ 
	F_2^{\rm sym}(\k_1,\k_2) &= \frac{-5k_{12,-}^{(2,2)}+3k_{12,+}^{(2,1)}s^2+2s^4}{28k_1^2k_2^2}\, ,\\
	G_2^{\rm sym}(\k_1,\k_2) &= -\frac{3k_{12,-}^{(2,2)}-k_{12,+}^{(2,1)}s^2+4s^4}{28k_1^2k_2^2}\, .
\end{align}
The dot products in $F_{\G_3}$ depends on more than one angle, so it is convenient to express it in terms of $s^2$ and the angle $\hat\k_2\cdot\hat\k_3$ so that we can directly apply our spin-exchange separable result \eqref{case3}. We have
\begin{align}
	(\hat\k_1\cdot\hat\k_2)(\hat\k_1\cdot\hat\k_3)(\hat\k_2\cdot\hat\k_3) &= \frac{(k_1^2+k_2^2-s^2)(2k_2k_3(\hat\k_2\cdot\hat\k_3)+k_3^2-k_4^2+s^2)(\hat\k_2\cdot\hat\k_3)}{4k_1^2k_2k_3}\, .
\end{align}
The $F_3^{\rm sym}$ term is rather complicated and depend on all of $s$, $t$, and $u$. However, as we described in the main text, it can be written in a manifestly crossing symmetric form in terms of $\hat F_3$ defined by
\begin{align}
	\hat F_3(\k_1,\k_2,\k_3) \equiv \frac{1}{9}+\frac{g_0(k_1,k_2,k_3,k_4)+g_s(k_1,k_2,k_3,k_4,s)}{3024(k_1k_2k_3)^2}\, ,\label{F3hat}
\end{align}
where
\begin{align}
	g_0(k_1,\cdots,k_4)  &\,\equiv\, -49\hs K^{(4,2)}+(24K^{(2,2)}-29K^{(4,0)})k_4^2-2K^{(2,0)}k_4^4\, ,\\[5pt]
	g_s(k_1,\cdots,k_4,s) & \,\equiv\, 3 k_{12,-}^{(1,2)}k_{12,+}^{(1,2)}k_{34,-}^{(1,1)} k_{34,+}^{(1,1)}(7k_3^2+2k_4^2)s^{-2}-3(7k_{12,+}^{(2,1)}-14k_3^2-2k_4^2)s^4-14s^6\nn[5pt]
	&\quad +\Big[7\big(5k_{12,-}^{(2,2)}+2k_{12,+}^{(2,1)}k_3^2-4k_3^4\big)+(23k_{12,+}^{(2,1)}+20k_3^2)k_4^2+8k_4^4\Big]s^2\, .
\end{align}
This basis is convenient since $F_3^{\rm sym}=\hat F_3^{\rm sym}$, but each permutation $\hat F_3$ depends on only one diagonal momentum, whereas $F_3$ depends on two.

\vskip 4pt 
Finally, let us list all the super-reduced trispectra associated with all combinations of bias parameters:
\begin{align}
	\tau_{2211}^{\delta^2}(\k_1,\k_2,\k_3,\k_4) &= 2b_\delta^2 b_{\delta^2}^2 D_1D_2^2D_3D_4^2 P(k_1)P(k_3)P(s)\, , \\
	\tau_{2211}^{\G_2}(\k_1,\k_2,\k_3,\k_4) &= 2b_\delta^2 b_{\delta^2}^2 D_1D_2^2D_3D_4^2 P(k_1)P(k_3)P(s) \Big[\sigma_{\k_1,-\k_{12}}^2\sigma_{\k_3,\k_{12}}^2\Big]\, ,\\
	\tau_{2211}^{\delta \times \delta^2}(\k_1,\k_2,\k_3,\k_4) &= 4b_\delta^3 b_{\delta^2} D_1D_2^2D_3D_4^2 P(k_1)P(k_3)P(s)\Big[F_2^{\rm sym}(\k_1,-\k_{12})\Big]\, , \label{tcross1}\\
	\tau_{2211}^{\delta \times \G_2}(\k_1,\k_2,\k_3,\k_4) &= 4b_\delta^3 b_{\G_2} D_1D_2^2D_3D_4^2 P(k_1)P(k_3)P(s)\Big[\sigma_{\k_1,-\k_{12}}^2\Big]\, , \label{tcross2}\\
	\tau_{2211}^{\delta^2 \times \G_2}(\k_1,\k_2,\k_3,\k_4) &= 4b_\delta^3 b_{\G_2} D_1D_2^2D_3D_4^2 P(k_1)P(k_3)P(s)\Big[F_2^{\rm sym}(\k_1,-\k_{12})\sigma_{\k_3,\k_{12}}^2\Big]\, ,\label{tcross3}
\end{align}
where $\delta\times \delta^2$, $\delta\times \G_2$, $\delta^2\times \G_2$ denote the cross-terms, and
\begin{align}
	\tau_{3111}^{\delta^2}(\k_1,\k_2,\k_3,\k_4) &= 2b_\delta^3 b_{\delta^2} D_1D_2D_3D_4^3\Big[ F_2^{\rm sym}(\k_1,\k_2) \Big]\, ,\\
	\tau_{3111}^{\delta^3}(\k_1,\k_2,\k_3,\k_4) &= b_\delta^3 b_{\delta_3} D_1D_2D_3D_4^3 \, ,\\
	\tau_{3111}^{\G_2}(\k_1,\k_2,\k_3,\k_4) &= 2b_\delta^3 b_{\G_2} D_1D_2D_3D_4^3 \Big[\sigma_{\k_{12},\k_3}^2 F_2^{\rm sym}(\k_1,\k_2)\Big] \, ,\\
	\tau_{3111}^{\G_2\delta}(\k_1,\k_2,\k_3,\k_4) &= b_\delta^3 b_{\G_2} D_1D_2D_3D_4^3 \Big[\sigma_{\k_1,\k_2}^2\Big] \, ,\\
	\tau_{3111}^{\G_3}(\k_1,\k_2,\k_3,\k_4) &= b_\delta^3 b_{\G_3} D_1D_2D_3D_4^3 \Big[\tfrac{3}{2}\sigma_{\k_1,\k_2}^2-(\hat\k_1\cdot\hat\k_2)(\hat\k_2\cdot\hat\k_3)(\hat\k_3\cdot\hat\k_1)\Big]\, ,\\
	\tau_{3111}^{\Gamma_3}(\k_1,\k_2,\k_3,\k_4) &= 2b_\delta^3 b_{\Gamma_3} D_1D_2D_3D_4^3 \sigma_{\k_{12},\k_3}^2\Big[F_2^{\rm sym}(\k_1,\k_2)-G_2^{\rm sym}(\k_1,\k_2)\Big]\,.
\end{align}
The total $s$-channel trispectrum can be obtained by summing over 8 permutations of each super-reduced trispectrum. The shapes of the corresponding angular trispectra were shown in \S\ref{sec:res}.

\section{Spin-Weighted Functions}\label{app:spin}

As we mentioned in the main text, there are alternative representations of separable trispectra in angular space. One such representation involves {\it spin-weighted harmonics}, whose origin we can understand as follows. When expressing correlation functions in three-dimensional space in terms the spherical coordinates, some momentum dependence can be traded with derivatives with respect to the angular coordinates. These derivatives, carrying directional information, in turn will transform scalars into spin-weighted fields on the sphere. For scalar correlators like we are studying, these spin weights necessarily cancel in the end and are thus fake, but still provides a useful way of describing angular observables. 
In this appendix, we present details on spin-weighted functions on a sphere and use these to express contact separable trispectra.

\subsection{Covariant Derivatives}

The first study of spin-weighted functions goes back to~\cite{Newman:1966ub, Goldberg:1966uu}, which we first briefly review. The line element of the three-dimensional Euclidean space in the spherical coordinates is
\begin{align}
	\d s^2 = \d r^2 + r^2\d\theta^2+r^2\sin^2\theta\hs\d\varphi^2\, .
\end{align}
It is convenient to work in terms of the orthonormal basis given by
\begin{align}
	 e_r &= \partial_r\, , \quad  e_\theta = \frac{1}{r}\,\partial_\theta \, , \quad  e_\varphi = \frac{1}{r\sin\theta}\,\partial_\varphi\, .
	 \end{align}
From these we can form a new, ``helicity'' basis vectors $e_\pm$ with their dual 1-forms $\omega^\pm$ by
\begin{align}
 e_\pm &= \frac{1}{\sqrt 2}( e_\theta \pm i  e_\varphi)\, ,\quad
\omega^\pm = \frac{1}{\sqrt 2}(r\hs\d\theta \mp ir\sin\theta\hs\d\varphi)\, .
\end{align}
These are defined such that under a standard rotation of angle $\gamma$ they transform as $e_\pm \to e^{\pm i\gamma}e_\pm$. 
The natural set of derivative operators we can use in this basis are the {\it spin-raising} operator and its adjoint, {\it spin-lowering} operator $\bar\eth$, whose action on a spin-$J$ function $_{J}f$ are defined as
\begin{align}
	\eth ({}_Jf) &= -(\partial_\theta +i\csc\theta\partial_\varphi + J\cot\theta){}_Jf\, ,\nn
	\bar\eth ({}_Jf)&= -(\partial_\theta -i\csc\theta\partial_\varphi - J\cot\theta){}_Jf\, .
\end{align}
These operators act very naturally on spherical harmonics, which we review at the end of this section. Standard Cartesian derivatives can be recast in terms of these two operators together with the radial derivative $\partial_r$. For example, given two scalar fields $f$ and $g$, we have
\begin{align}
	\partial^a f\partial_a g 
	&=\partial_r f \partial_r g+\frac{1}{2r^2}(\eth f\bar\eth g+\bar\eth f\eth g)\, ,
\end{align}
where $\partial_\pm f\equiv e_\pm(f)$. Going beyond first derivatives requires us to work out covariant derivatives, which in the orthonormal basis are defined in terms of the connection 1-form ${\omega^a}_b$ by the relation $\nabla_{ e_a} e_b = ({\omega^c}_a)_b e_c$. To obtain these, we first compute the exterior derivatives of the dual 1-forms
\begin{align}
	\d \omega^0 &= 0\, ,\quad
	\d \omega^\pm
	= \frac{1}{r}\omega^0 \wedge \omega^\pm - \frac{\cot\theta }{\sqrt 2}\omega^\pm\wedge \omega^\mp\, .
\end{align}
Comparing these exterior derivatives with Cartan's first structure formula, we deduce that
\begin{align}
	\begin{array}{l} \d \omega^0 = -{\omega^0}_a\wedge \omega^a \\[5pt]  \d \omega^\pm = -{\omega^\pm}_a\wedge \omega^a \end{array}\quad \Rightarrow \quad {\omega^\pm}_0 = \frac{1}{r}\, \omega^\pm\, , \quad\, {\omega^\pm}_\mp = \frac{\cot\theta}{\sqrt 2\hs r}\,\omega^\pm\, .
\end{align}
Using these, we can express double covariant derivatives in terms of $\{\eth,\bar\eth,\partial_r\}$ as
\begin{align}
	\nabla_r\nabla^r f &= \partial_r^2 f\, ,\quad \nabla_+\nabla^r f = \frac{1}{\sqrt 2 r^2}\eth(r\partial_r f-f)\,,\quad	\nabla_+\nabla^+ f = \frac{1}{2r^2}(\eth\bar\eth + 2r\partial_r) f\, ,
\end{align}
and similarly for $\nabla_-$. For example, some relevant formulas for their action on scalar fields are
\begin{align}
	\nabla_a\nabla^a f &=\frac{1}{r^2}\partial_r(r^2\partial_r f) + \frac{1}{2r}(\eth\bar\eth +\bar\eth\eth )f\, ,\label{grad1}\\
	\nabla_a\nabla_b f\nabla^a\nabla^b g &=\frac{1}{4r^4}\Big[(\eth\eth f)(\bar\eth\bar\eth g)+\left(\eth\bar\eth f+2r\partial_r f\right)\left(\bar\eth\eth g+2r\partial_r g\right)\nn
	&\qquad +4(r\partial_r \eth f-\eth f)(r\partial_r \bar\eth g-\bar\eth g) + c.c.\Big]+(\partial_r^2f)(\partial_r^2g)\, .\label{grad2}
\end{align}
We show the role played by these derivatives in computing certain angular correlations in \S\ref{app:angtri}.

\subsubsection*{Spin-Weighted Spherical Harmonics}
Spin-$j$ spherical harmonic ${}_jY_{\ell m}$ is defined by the action of the spin-raising operator on the usual spherical harmonic $Y_{\ell m}$ as
\begin{align}
	\eth ({}_j Y_{\ell m}) = \sqrt{(\ell-j)(\ell+j+1)}\, {}_{j+1} Y_{\ell m}\, ,\\
		\bar\eth ({}_j Y_{\ell m}) = -\sqrt{(\ell+j)(\ell-j+1)}\, {}_{j-1} Y_{\ell m}\, ,
\end{align}
with ${}_j Y_{\ell m} = \sqrt{(\ell-j)!/(\ell+j)!}\,\eth^jY_{\ell m}$ and $_{j}Y_{\ell m}=0$ if $|j|>\ell$. Their explicit representation in terms of the angles $\theta$ and $\varphi$ is
\begin{align}
	{}_jY_{\ell m}(\theta,\varphi) &=\sqrt{\frac{(\ell+m)!(\ell-m)!}{(\ell+j)!(\ell-j)!}\frac{2\ell+1}{4\pi}} \sin^{2\ell}\tfrac{\theta}{2}\nn
	&\times \sum_r\begin{pmatrix}
		\ell-j \\ r
	\end{pmatrix}\begin{pmatrix}
		\ell+j \\ r+j-m
	\end{pmatrix}(-1)^{\ell-r-j}e^{im\varphi}\cot^{2r+j-m}\tfrac{\theta}{2}\, .
\end{align}
The tensor spherical harmonics have the useful property that they are orthonormal functions on the sphere
\begin{align}
	\int_{S^2}\d\Omega_{\hat\n}\, {}_j Y_{\ell m}(\hat \n){}_j Y_{\ell' m'}^*(\hat \n) = \delta_{\ell\ell'}\delta_{mm'}\, .
\end{align}
This together with the identity that relates a product of two spherical harmonics to a sum over spherical harmonics is
\begin{equation}
\hskip -7pt	{}_{j_1}\!Y_{\ell_1 m_1}{}_{j_2}\!Y_{\ell_2 m_2} = \sum_{\ell_3m_3} \!\sqrt{\frac{(2\ell_1+1)(2\ell_2+2)(2\ell_3+3)}{4\pi}}\!\begin{pmatrix}
			\ell_1 & \ell_2 & \ell_3 \\ -j_1 & -j_2 & -j_3 \end{pmatrix}\!\!\begin{pmatrix}
			\ell_1 & \ell_2 & \ell_3 \\ m_1 & m_2 & m_3 \end{pmatrix}{ {}_{j_3}Y_{\ell_3 m_3}^*}\hs ,
\end{equation}
allows us to easily perform angular integration involving these functions; for example, the integral over three spin-weighted spherical harmonics is
\begin{align}
		&\int_{S^2}\d\Omega_{\hat\n}\, {}_{j_1} Y_{\ell_1 m_1}(\hat \n){}_{j_2} Y_{\ell_2 m_2}(\hat \n){}_{j_3} Y_{\ell_3 m_3}(\hat \n) \nn
		&\qquad\qquad=\sqrt{\frac{(2\ell_1+1)(2\ell_2+1)(2\ell_3+1)}{4\pi}}\begin{pmatrix}
			\ell_1 & \ell_2 & \ell_3 \\ -j_1 & -j_2 & -j_3 \end{pmatrix}\begin{pmatrix}
			\ell_1 & \ell_2 & \ell_3 \\ m_1 & m_2 & m_3 \end{pmatrix}\,.
\end{align}
For $j_1=j_2=j_3=0$, this reduces to the Gaunt integral \eqref{gaunt}.

\subsection{Angular Trispectrum}\label{app:angtri}
Let us now see how a contact separable trispectrum can be represented in angular space in terms of spin-weighted spherical harmonics. We consider a trispectrum that depends on $(\k_3\cdot\k_4)^{J}$ in the $s$-channel, which in momentum space takes the form
\begin{align}
	\langle \O_1\cdots \O_4\rangle &=f_1(k_1,z_1)\cdots f_4(k_4,z_4)(\k_3\cdot\k_4)^{J}\times (2\pi)^3 \Ddelta(\k_1+\k_2+\k_3+\k_4) \\
	&=f_1(k_1,z_1)\cdots f_4(k_4,z_4) \int_{\mathbb{R}^3}\d^3r\, e^{i\k_1\cdot\r}e^{i\k_2\cdot\r}\nabla_{a_1}\cdots \nabla_{a_{J}} e^{i\k_3\cdot\r}\nabla^{a_1}\cdots \nabla^{a_{J}}e^{i\k_4\cdot\r}\, ,\nonumber
\end{align}
where we have traded the dot products with gradients acting on the appropriate plane waves. Expanding the planes waves in spherical harmonics gives, these gradients turn into radial and angular derivatives acting on the spherical Bessel functions and spherical harmonics, respectively. 
To illustrate how this works in practice, let us explicitly work out the case $J=1$. Using the expansion \eqref{eq:rayleigh} and then performing the angular integrations, we get
\begin{align}
	\langle  \O_1\cdots \O_4\rangle &=\sum_{\ell_1'm_1'}\cdots \sum_{\ell_4'm_4'} \prod_{i=1}^4\Big[f_i(k_i,z_i)Y_{\ell_i'm_i'}(\hat\k_i)\Big] \int_0^\infty\d r\, j_{\ell_1'}(k_1r)j_{\ell_2'}(k_2r)\int_{S^2}\d\Omega_{\hat\n} Y_{\ell_1'm_1'}(\hat\r)Y_{\ell_2'm_2'}(\hat\r)\nn
	&\hskip-60pt {\times} \bigg[r^2\partial_r j_r(k_3r)\partial_r j_r(k_4r) Y_{\ell_3'm_3'}(\hat\r)Y_{\ell_4'm_4'}(\hat\r){+}\frac{j_r(k_3r)j_r(k_4r)}{2}(\eth Y_{\ell_3'm_3'}(\hat\r)\bar\eth Y_{\ell_4'm_4'}(\hat\r){+}c.c.)\bigg] .
\end{align}
Plugging this into the projection formula \eqref{angnpt} and performing the angular integrations, we obtain the reduced trispectrum (for $J=1$)
\begin{align}
	t^{\ell_1\ell_2}_{\ell_3\ell_4}(L) &=  \frac{h^{\ell_1\ell_2L}h^{\ell_3\ell_4L}}{(2\pi^2)^4}\int_0^\infty\d r \,r^2 I_{\ell_1}^{(1)}(r)I_{\ell_2}^{(2)}(r)\partial_r I_{\ell_3}^{(3)}(r)\partial_r I_{\ell_4}^{(4)}(r)\nn[3pt]
	&+\frac{h^{\ell_1\ell_2L}(h^{\ell_3\ell_4L}_{-110}+h^{\ell_3\ell_4L}_{1-10})}{2(2\pi^2)^4}\int_0^\infty\d r\, I_{\ell_1}^{(1)}(r)I_{\ell_2}^{(2)}(r)I_{\ell_3}^{(3)}(r) I_{\ell_4}^{(4)}(r)\, ,
\end{align}
where we have defined
\begin{align}
	h^{\ell_1\ell_2\ell_3}_{j_1j_2j_3} \equiv h^{\ell_1\ell_2\ell_3}\prod_{i=1}^3\big[(-1)^{(|j_i|-j_i)}(\ell_i-j_i)_{j_i}(\ell_i+1)_{j_i}\big]^{1/2} \,,
\end{align}
and $(a)_n\equiv \Gamma(a+n)/\Gamma(a)$ is the Pochhammer symbol. It is straightforward to use \eqref{grad1} and \eqref{grad2} to similarly work out the formula for $J=2$.

\newpage

\newpage
\bibliographystyle{utphys}
\bibliography{Trispectrum}

\end{document}